\documentclass[a4paper,fleqn,usenatbib]{mnras}

\usepackage{newtxtext,newtxmath}

\usepackage[T1]{fontenc}
\usepackage{ae,aecompl}

\usepackage{graphicx}	
\usepackage{amsmath}	
\usepackage{amssymb}	

\usepackage{xcolor}
\usepackage{tablefootnote}
\usepackage{supertabular}
\makeatletter
\newcommand\footnoteref[1]{\protected@xdef\@thefnmark{\ref{#1}}\@footnotemark}
\newcommand\tablefootnoteref[1]{\protected@xdef\@thefnmark{\ref{#1}}\@tablefootnotemark}
\makeatother
\usepackage{array}
\usepackage{multirow}
\usepackage[above]{placeins}
\usepackage{float}

\title[Simulated emission from the CGM]{Emission from the circumgalactic medium: from cosmological zoom-in simulations to multiwavelength observables}
\author[R. Augustin et al.]{R. Augustin$^{1,2}$\thanks{E-mail: ramona.augustin@lam.fr}, 
S. Quiret$^{1}$,
B. Milliard$^{1}$, 
C. P\'eroux$^{1}$, 
D. Vibert$^{1}$, 
J. Blaizot$^{3}$,
\newauthor
Y. Rasera$^{4}$, 
R. Teyssier$^{5}$,
S. Frank$^{6}$,
J.-M. Deharveng$^{1}$,
V. Picouet$^{1}$,
D. C. Martin$^{7}$,
\newauthor
E. T. Hamden$^{7,8}$,
N. Thatte$^{9}$,
M. Pereira Santaella$^{9}$,
L. Routledge$^{9}$,
S. Zieleniewski$^{9,10}$\\\\\\
$^{1}$Aix Marseille Universit\'e, CNRS, LAM (Laboratoire d'Astrophysique de Marseille) UMR 7326, 13388, Marseille, France\\
$^{2}$European Southern Observatory, Karl-Schwarzschildstrasse 2, D-85748 Garching bei M{\"u}nchen, Germany\\
$^{3}$Univ Lyon, Univ Lyon1, Ens de Lyon, CNRS, Centre de Recherche Astrophysique de Lyon UMR5574, F-69230, Saint-Genis-Laval, France\\
$^{4}$LUTH, Observatoire de Paris, PSL Research University, CNRS, Universit\'e Paris Diderot, Sorbonne Paris Cit\'e,\\ 5 Place Jules Janssen, 92195 Meudon, France\\
$^{5}$Institute for Computational Science, University of Z{\"u}rich, Winterthurerstrasse 190, 8057 Z{\"u}rich, Switzerland\\
$^{6}$The Ohio State University, Department of Astronomy, Columbus, Ohio 43210, USA\\
$^{7}$California Institute of Technology, MC 405-47, 1200 East California Boulevard, Pasadena, CA 91125, USA\\
$^{8}$University of Arizona, Steward Observatory, 933 N Cherry Ave, Tucson, AZ 85721, USA\\
$^{9}$Department of Physics, University of Oxford, Denys Wilkinson Building, Keble Road, Oxford OX1 3RH, UK\\
$^{10}$King's College School, Southside, Wimbledon Common, London, SW19 4TT, UK
}

\begin{document}
\def\mean#1{\left< #1 \right>}


\pagerange{\pageref{firstpage}--\pageref{lastpage}} \pubyear{2019}

\maketitle

\label{firstpage}

\begin{abstract}

We simulate the flux emitted from galaxy halos in order to {quantify the} brightness of the circumgalactic medium (CGM).
We use dedicated zoom-in cosmological simulations with the hydrodynamical Adaptive Mesh Refinement code RAMSES, which are evolved down to z=0 and reach a maximum spatial resolution of 380 $h^{-1}$pc and a gas mass resolution up to 1.8$\times 10^{5} h^{-1} \rm{M}_{\odot}$ in the densest regions.
We compute the expected emission from the gas in the CGM using CLOUDY emissivity models for different lines (e.g. Ly$\alpha$, CIV, OVI, {CVI, OVIII}) considering UV background fluorescence, gravitational cooling and continuum emission.
In the case of Ly$\alpha$ we {additionally} consider the scattering of continuum photons.
We compare our predictions to current observations and find them to be in good agreement at any redshift {after adjusting the Ly$\alpha$ escape fraction}.
{We combine our mock observations with instrument models for FIREBall-2 (UV balloon spectrograph) and HARMONI (visible and NIR IFU on the ELT) to predict CGM observations with either instrument and optimise target selections and observing strategies.
Our results show that Ly$\alpha$ emission from the CGM at a redshift of 0.7 will be observable with FIREBall-2 for bright galaxies (NUV$\sim$18 mag), while metal lines like OVI and CIV will remain challenging to detect.
HARMONI is found to be well suited to study the CGM at different redshifts with various tracers.
}

\end{abstract}


\section{Introduction}

Understanding the complex mechanisms regulating galaxy formation is one of the main questions today in cosmology and astrophysics. 
The question of how galaxies gather gas to sustain star formation is of particular interest, as it could shed light on the fact that the star formation rate (SFR) has been declining from $\rm z\sim2$ while diffusely distributed hydrogen still is the dominant component for the total baryonic mass budget (as compared to hydrogen in stars, \citealt{Madau2014}).
Numerical simulations bring valuable insight into accretion mechanisms that replenish the gas reservoir of star formation. 
The two main mechanisms are cold accretion from dense flows of cold gas, and the hot accretion of more diffuse gas from the halo. Due to the current scarcity of direct observations, these are vividly debated (e.g \citealt{Keres2005,Dekel2009,Bournaud2011,Fox2017}). 
Simultaneously, powerful gas outflows provide negative feedback on star formation.
These outflows have been observed with various techniques and instruments, but their numerical implementation remains challenging
\citep{Pettini2001,Steidel2010,Vogelsberger2014}.

The circumgalactic medium (CGM) of galaxies, at the interface between galaxies and the intergalactic medium (IGM), is loosely defined as the region within {$\sim$}300 kpc \citep{Steidel2010,Shull2014,Tumlinson2017} where these outflowing and accreting mechanisms are interacting. 
Studying the CGM will {provide} key constraints on the question of galaxy formation and evolution. 
Absorption spectroscopy has already {shed light} on the distribution and the chemical composition of the CGM gas, but only on a statistical point of view, given that only one line of sight per galaxy can be probed due to the scarcity of background quasars in the vicinity of galaxies
\citep{Noterdaeme2012,Pieri2014,Quiret2016,Rahmani2016,Krogager2017,Augustin2018}.
\citet{Hummels2017} have implemented a technique to create mock absorption spectra from cosmological simulations in order to understand the gas we see in absorption, yet mapping the emission of the CGM is the natural next step to fully understanding the CGM.
Its low surface brightness makes direct observation challenging, but there has been tremendous progress over the last years in order to find faint emission around galaxies.
At high redshifts, large ground based telescopes such as the Very Large Telescope (VLT), Subaru or Keck offer the first hints of Ly$\alpha$ emission CGM mapping, achieved through the stacking of a large number of systems \citep{Steidel2011,Momose2014}, long exposures \citep{Rauch2008, Wisotzki2016,Wisotzki2018} or by selecting objects whose Ly$\alpha$ luminosity is boosted by the presence of a bright quasar nearby \citep{Cantalupo2014, Martin2014a,Borisova2016,fab15,ArrigoniBattaia2016,ArrigoniBattaia2018,Arrigoni2018qmi}. 
\citet{Gronke2017} have shown that indeed the findings so far agree with the Ly$\alpha$ profiles found from simulations and extended low surface brightness gas around galaxies.
{Using narrow band imaging with HST, \citet{Hayes2016} have discovered also extended OVI emission around a z=0.2 galaxy and thereby created one of the first metal line maps of the CGM.}

Understanding the different processes responsible for diffuse emission in this region is of great interest. 
An accurate modelling of these processes would require spatial resolution of a few parsecs and accounting for dust effects and {radiative} transfer, which is only currently achievable for single galaxy simulations (e.g. \citealt{Rosdahl2012,Oppenheimer2016}). 
At the same time, the study of co-evolution between galaxies and the IGM has to be conducted on much larger scales. 
Moreover, modelling the emission from this medium requires chemical-photo-ionization calculations, that are simply too heavy to be produced on the fly. 
Post-processing radiative transfer analysis is so far the best tool to estimate a realistic level of emission.
\citet{Bertone2012} have used the hydrodynamic cosmological simulation {OWLS \citep{Schaye2010}} to analyse the strength of UV lines in the CGM and predicted the brightest emission line to be from \ion{H}{i} Ly$\alpha$ (1216 \AA) and the strongest metal line to be \ion{C}{iii} (977 \AA).
\citet{silva16} have analytically studied the emission of Ly$\alpha$ at $z$<3 in filaments and their detectability.
They found that future space based experiments with a sensitivity  of 3.7$\times$10$^{-9}$ erg/s/cm$^2$/sr ($\sim$37 mag/arcsec$^2$) will be able to detect hydrogen and helium in intergalactic filaments.
{\cite{Lokhorst2019} have used EAGLE simulations with a CLOUDY emission model to predict the fluxes from the CGM and IGM and investigated the detectability of faint H$\alpha$ emission at low redshifts. They found that the Dragonfly Telephoto Array\footnote{http://www.dragonflytelescope.org/}, equipped with suitable narrowband filters, would be able to directly map the cosmic web. 
}

{Very recently, a number of cosmological simulations have highlighted the importance of increased resolution in the CGM in the context of making predictions for observations. 
Works by \cite{Hummels2018} and \cite{Suresh2019} have investigated the importance of highly resolved CGM in simulations in order to reproduce the observed cool gas column densities.
Similarly, \cite{vandeVoort2019} have found that the covering fraction of cool gas as traced by HI absorption increases dramatically with increasing spatial resolution.
\citet{Corlies2018} and \citet{Peeples2019} have presented new Enzo AMR simulations with forced refinement in the CGM and made predictions for both absorption line studies as well as emission line maps as seen with current IFUs, confirming the resolution effect on CGM studies.
}

Here, we present a new simulation run of RAMSES \citep{Teyssier2002} over a box of 100 comoving Mpc/h with a zoom in over a region of 13.92 comoving Mpc/h and our post-processing of snapshots to obtain flux maps and 3D data cubes of individual galaxies and their CGM. 
These are used to predict the expected flux of different lines to enable comparison with observations.
{We then use those 3D data cubes to create mock observations of the CGM with two different instruments: FIREBall-2 and HARMONI.}

This work is structured as follows: 
In section 2 we present our cosmological zoom-in simulations, in section 3 the photo-ionization model we are using and in section 4 the {comparison to observations. 
In section 5 we will use the simulated halos to prepare FIREBall-2 target selection and data analysis. 
We assess the compatibility of ELT/HARMONI for CGM observations in section 6.
Our conclusions are given in section 7}.
We assume a flat $\Lambda$CDM universe with $\Omega_{\Lambda}$ = 0.742, $\Omega_{m}$ = 0.258, $\Omega_{b}$ = 0.045, H0 = 71.9, $\sigma_{8}$=0.798, $\rm n_{s}$=0.963.

\section{Cosmological zoom-in simulations}

\begin{table*}
\caption{Initial parameters for the zoom simulation. {The parameters are analogous to \citet{Teyssier2013} and adjusted to the resolution in our zoom-in simulation.}
}
\label{tab:ramsesparameters}
\begin{tabular}{ccc}
\hline
Parameter & Description & Value \\ \hline
$\rm \epsilon_{*}$ & star formation efficiency & 0.01 \\
$\rm n_{*}$ & star formation density threshold in H/cc & 3 \\
$\rm T2_{*}$ & ISM polytropic temperature in K/$\mu$& 3000 \\
$\rm \eta_{sn}$ & supernova mass fraction & 0.2 \\
yield & supernova metal yield & 0.1 \\
$\rm mass_{GMC}$ & stochastic exploding GMC mass in solar mass & 2$\rm \times 10^{6}$ \\
$\rm z_{reion}$ & reionization redshift & 20 \\
$\rm \Omega_{m}$ & matter density & 0.26 \\
$\rm \Omega_{l}$ & vacuum density & 0.74 \\
$\rm \Omega_{b}$ & baryonic matter density & 0.045 \\
$\rm \Omega_{k}$ & spatial curvature density & 0 \\
$\rm H_{0}$ & Hubble parameter in km/s & 71.9 \\
$\rm \sigma_{8}$ & amplitude of the (linear) power spectrum on the scale of 8 $h^{-1}$ Mpc & 0.798 \\
$\rm n_{s}$ & primordial spectral index of scalar fluctuations & 0.963 \\
\hline
\end{tabular}
\end{table*}

\begin{table*}
\caption{Comparison of our zoom simulation with the low resolution simulations of \citet{Frank2012}.}
\label{tab:frankcomparison}
\begin{tabular}{ccc}
\hline
 \citet{Frank2012}  & Parameter &  zoom-in (this work) \\ 
 ``low-resolution" simulation &  & ``high-resolution" simulation \\ \hline
1.53 kpc $h^{-1}$ & maximum spatial resolution & 0.38 kpc $h^{-1}$ \\
$\sim$4.42$\rm \times 10^{8} ~ M_{\odot}$ $h^{-1}$ & maximum mass resolution for dark matter &$\sim$8.7$\rm \times 10^{5} ~ M_{\odot}$ $h^{-1}$ \\
 $\sim$134$\rm \times 10^{6}$ & number of dark matter particles & $\sim$205$\rm \times 10^{6}$ \\
  $\rm 512^{3}$ & number of initial gas cells & $\rm 128^{3}$ \\
 100 comoving Mpc $h^{-1}$ & box length & 13.92 comoving Mpc $h^{-1}$  \\
 7 & maximum level of refinement & 18  \\
\hline
\end{tabular}
\end{table*}

We use cosmological simulations that were produced with the RAMSES \citep{Teyssier2002} grid-based hydrodynamical solver with adaptive mesh refinement (AMR) using $\sim$ 1.3 million CPU hours.
The basis of this work is presented in \citet{Frank2012}.
They have used cosmological simulations and CLOUDY modelling to predict the line fluxes of three UV lines (Lya at 1216 \AA , OVI at 1032/1038 \AA\ and CIV at 1548/1551 \AA ).
They found that the CGM is expected to have high enough emission (e.g. $\rm log(L_{Ly\alpha}) \sim 40.9-41.8$ erg/s) to be detectable with upcoming instruments but were less optimistic for filament detections from the IGM.

We aim to follow their approach in predicting CGM luminosities.
However, many physical processes in gas clumps in and around galaxies are taking place on scales that are lower than what can be resolved in the simulations.
With this in mind, we picked the most massive halo from the \citet{Frank2012} simulations {(which is also the most luminous in Ly$\alpha$ in their simulation)} and performed a zoom-in on the region around this halo.
{The halo was selected because of its high mass, which results in high density gas cells.
Indeed, in the AMR framework, the densest regions have the highest spatial resolution. 
This high spatial resolution allows us to distinguish between CGM and ISM and provides the basis for a detailed CGM study.
The high mass of the halo} could introduce a caveat towards low redshift (z $\lesssim$1) where this halo might not be representative of the average population. 
It corresponds to a massive group of galaxies rather than an isolated galaxy.
However, at low redshifts, galaxies typically exist in groups and clusters, and thus the simulation will probe the CGM in realistic environments.
We use the MUlti-Scale Initial Conditions code (MUSIC, \citealt{Hahn2013}) to zoom on a large cubic region with a box length of 13.92 Mpc/h. 
The simulation was performed using non thermal supernova (SN) feedback \citep{Teyssier2013} and '\textit{on-the-fly}' self-shielding. 
The latter {disables} the ionizing background for cells with a neutral hydrogen density $\rm n_{HI}>0.01 at/cc$. 
This reproduces the self-shielding of gas cells in dense regions from ionizing background radiation and gives a good prediction of the temperature in the absence of radiative transfer. 
The threshold value is based on radiative transfer studies that have derived an estimate on the density at which the fraction of neutral hydrogen becomes dominant \citep{Faucher-Giguere2010,Rosdahl2012}.
The maximum refinement level is set to 18, giving a spatial resolution in the densest region of the simulation of about 380 $h^{-1}$ comoving parsecs and a typical resolution of 1-2 comoving kpc  $h^{-1}$ in the CGM regions. 
A list of parameters describing the initial conditions {(adapted from \cite{Teyssier2013} to the resolution in our simulations)} of our simulations is given in table \ref{tab:ramsesparameters} and a comparison with the previous low-resolution simulation is given in table \ref{tab:frankcomparison}).
The final number of particles is $\rm \sim$ 205 million for Dark Matter, 51 million for stars and 592 million gas cells, and at $\rm z=0$, the central zoomed halo mass is about $\rm 3 \times 10^{13}~h^{-1}~M_{\odot}$, with $\rm \sim$ 30 million particles. 
This makes it one of the largest DM+hydro+SF zoom simulations to date.
Our analysis follows that of \citet{Frank2012} and \citet{Bertone2012} with an increased resolution enabling us to probe colder and denser gas. 

\begin{figure*}
\centering
\includegraphics[width=.329\textwidth]{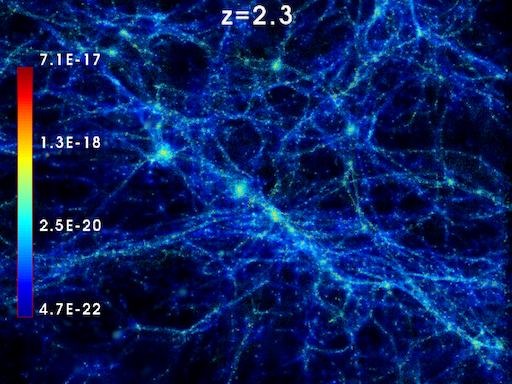}
\includegraphics[width=.329\textwidth]{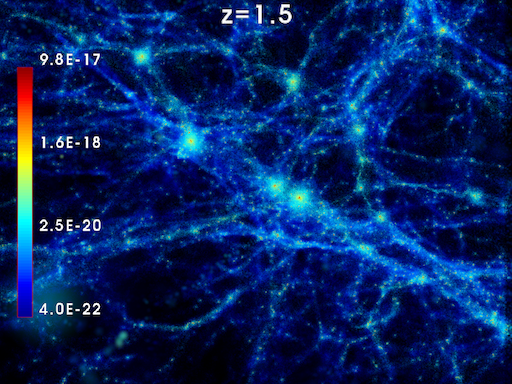}
\includegraphics[width=.329\textwidth]{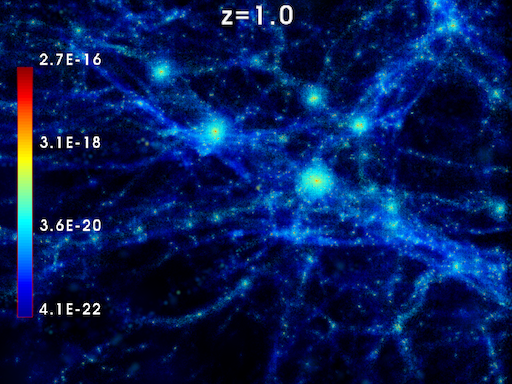}
\includegraphics[width=.329\textwidth]{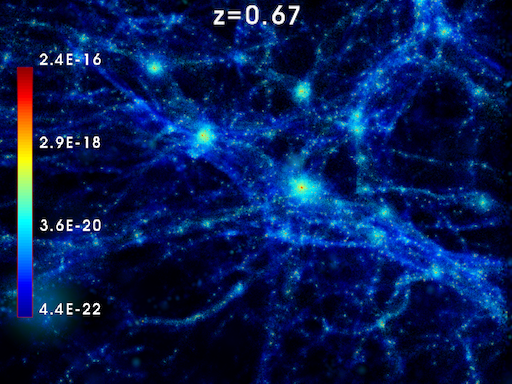}
\includegraphics[width=.329\textwidth]{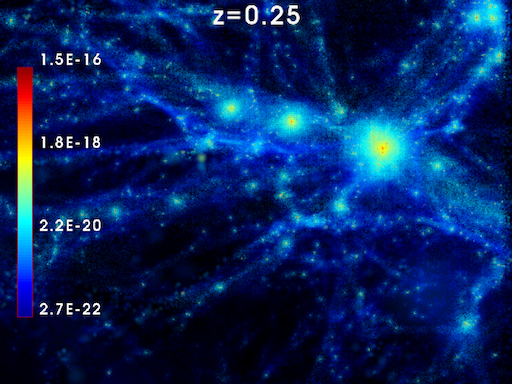}
\includegraphics[width=.329\textwidth]{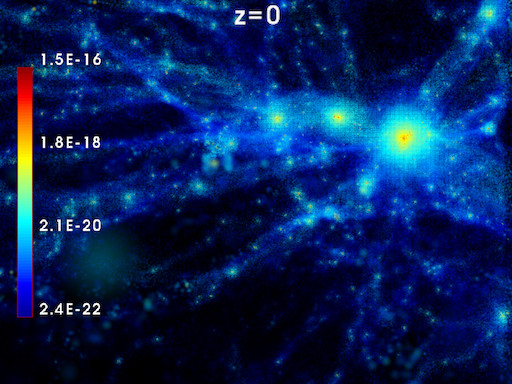}
\caption{Evolution of the gas distribution in the high-resolution simulation for 6 different redshifts: z=2.3, 1.5, 1.0, 0.67, 0.25, 0 (from top left to bottom right). Each panel corresponds to the `zoomed' region of the high-resolution simulation, which is about 13.5 comoving Mpc/h. The filamentary structure of the IGM, and the formation of the massive halos are apparent. The color bar indicates the gas density in units of $\rm kg/m^{3}$.}
\label{im:snaphsotevolution}
\end{figure*}

Fig. \ref{im:snaphsotevolution} shows six snapshots of the zoom-in region of the bright halo at different redshifts ($\rm z=2.33,~1.5,~1.0,~0.67,~0.25,~0.0$ respectively). We can see the progressive formation of the {most massive} halo from $\rm z=1.0$.
The web-like structure of the IGM clearly emerges in each of these snapshots, where we see faint filaments connecting overdense regions. 
We also witness the presence of isolated halos within each filament.
The properties of the {most massive} halo at different redshifts are gathered in table \ref{tab:haloprops}.
In this table we only go down to a redshift of 0.25 because the {processing of the halos and calculation of those values} for the z=0 snapshot take comparatively long and the data are not used in any later analysis.

\begin{table*}
\centering
\caption{{Most massive} halo characteristics for different redshifts. 
We give the redshift in the first column and the corresponding age of the Universe in the second. 
Columns 3-5 give the dark matter, stellar and gas mass within the virial radius, which is given in physical as well as comoving units in columns 6 and 7.
Column 8 provides the star formation rate.
}
\label{tab:haloprops}
\begin{tabular}{cccccccc}
\hline\hline
z & $\rm t_{Universe}$ & $\rm M_{DM}$ & $\rm M_{\star}$ & $\rm M_{gas}$ & $\rm R_{vir}$ & $\rm R_{vir}$ & SFR \\ 
 & [$\rm Gyr$] & [$\rm 10^{12} M_{\odot}$] & [$\rm 10^{12} M_{\odot}$] & [$\rm 10^{12} M_{\odot}$] & [$\rm pkpc$] & [$\rm ckpc$] & [$\rm M_{\odot}/yr$] \\\hline
 9.0 & $\rm \sim 0.6$ & $\rm 0.054$ & $\rm 0.00015 $  & $\rm 0.005$ & $\rm 9$ & $\rm 94$ & $\rm 2$ \\
 4.0 & $\rm \sim 1.6$ & $\rm 0.6$ & $\rm 0.013 $  & $\rm 0.09$ & $\rm 55$ & $\rm 274$ & $\rm 95$ \\
 2.3 & $\rm \sim2.9$ & $\rm 2.4$ & $\rm 0.122 $  & $\rm 0.26$ & $\rm 128$ & $\rm 423$ & $\rm 351$ \\
 1.0 & $\rm \sim 5.9$ & $\rm 9.6 $ & $\rm 0.852$ & $\rm 1.36$ & $\rm 327$ & $\rm 655$ & $\rm 622$ \\
 0.67 & $\rm \sim 7.5$ & $\rm 13.6 $ & $\rm 1.11$ & $\rm 1.61$ & $\rm 391$ & $\rm 652$ & $\rm 246$ \\
 0.25 & $\rm \sim 10.8$ & $\rm 28.3 $ & $\rm 3.37$ & $\rm 4.0$ & $\rm 679$ & $\rm 848$ & $\rm 333$ \\
\hline
\end{tabular}
\end{table*}

\begin{figure}
\centering
\includegraphics[width=.5\textwidth]{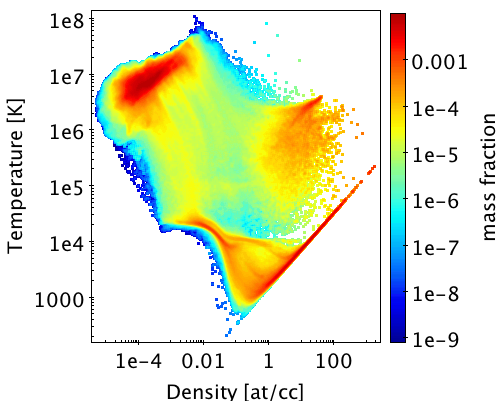}
\caption{2D histogram of the density-temperature phase diagram of the selected {most massive} halo of our zoom simulation at z=0.67.
{The density is given as total hydrogen density in at/cc.}
The color indicates the mass fraction of each bin (density bin width: 0.046 dex, temperature bin width: 0.054 dex). 
We find most of the gas in the IGM region (high temperatures, low densities).
} 
\label{im:nhitdiagram}
\end{figure}

\begin{figure}
\centering
\includegraphics[width=.5\textwidth]{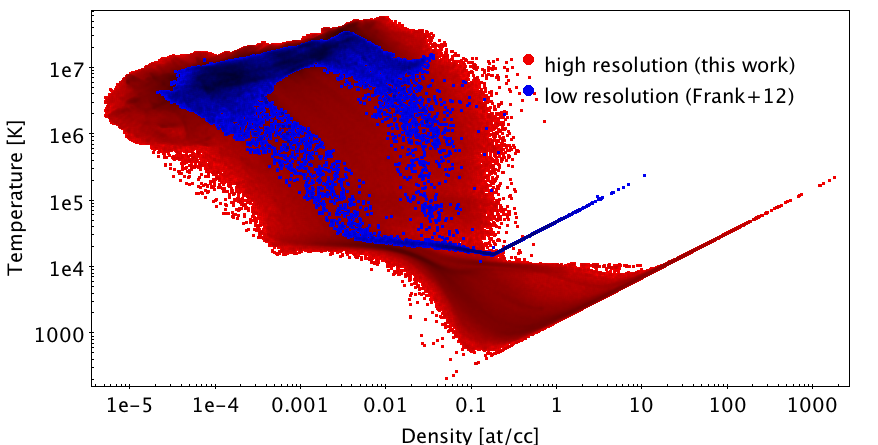}
\caption{Density-temperature phase diagram of selected halo for low-resolution (up to 1.53 kpc/h, blue) and high-resolution simulation (0.38 kpc/h, red).
{The density is given as total hydrogen density in at/cc.}
In our high resolution simulation we can now reach lower temperatures and higher densities compared to the low resolution simulations from \citet{Frank2012}.
} 
\label{im:densitytempdiagramzoomedhalo}
\end{figure}

Fig. \ref{im:nhitdiagram} shows the density temperature diagram for the zoomed halo. 
The colors indicate the mass fraction of each bin in the 2D histogram.
We find most of the gas to be in the IGM (low density, high temperature) region, where also most of the gas cells lie.
There is a second, smaller peak of gas cells around the so-called 'gutter' around T$\sim 10^{4}$K, where the cooling processes are in equilibrium with heating from external sources, which also holds a significant amount of gas mass.
Most of the rest of the mass is in the ISM {(n > 3 at/cc)}, although due to the resolution limit, these are concentrated in very few cells.

Fig. \ref{im:densitytempdiagramzoomedhalo} shows the same diagram in red (but with the delayed-cooling cells already taken away, see section 3.1.3) and the same halo in the low-resolution (blue points) simulation.
{
As expected, the high-resolution simulation extends to a larger parameter space, due to the higher number of cells in total in the halo.
} 
Two striking differences between these two simulations lie at the high density and low temperature end of Fig. \ref{im:densitytempdiagramzoomedhalo}. 
First, as the resolution increases, gas with higher densities can be sampled near the centre of the gravitational well.
{
This gas represents the high density regions in the ISM.
}
Consequently, the density threshold for star formation has been increased from $\rm n_{H}=0.1 at/cc$ \citep{Frank2012} to $\rm n_{H}=3 at/cc$ at a star formation efficiency of 1\%.
{The cells following a polytropic floor in both simulations are an artifact from the simulation code to artificially stabilize the gas versus the Jeans criterion at the resolution limit.}

{
The self-shielding of the gas leads to further gas cooling in the high-resolution simulation.
The coupling of this '\textit{on-the-fly}' self-shielding option with the effect of the significantly higher resolution results in the emergence of {dense and cool} gas phase, for which gas cells reach temperatures below $\rm 10^{4}K$, with densities higher than $\rm 0.1 at/cc$.
Such low temperatures could not be reached in the previous simulation set \citep{Frank2012}.
These cells are clearly identified in the visual inspection of the simulation through the presence of discs. 
{Examples are shown in Figure \ref{fig:discs} in the Appendix.}
Here, the simulation reaches its limits as the spatial resolution in the high density zones is of $\rm 381 pc/h$ (comoving).
Since we are not trying to resolve the ISM but are focused on the circumgalactic medium, we {conclude we reached the necessary limit in resolution where we can distinguish between galactic discs and the CGM}.
}

Another addition to the zoom simulation is the implementation of non-thermal supernova (SN) feedback from \citet{Teyssier2013}.
In hydrodynamical cosmological simulations there are typically two ways to simulate the feedback from supernovae or AGN activity: the momentum-driven feedback and the energy-driven feedback \citep{Costa2014}. 
The former injects pressure to the neighboring gas cells of a star particle undergoing supernova, acting a bit like a 'velocity kick', while the latter directly injects thermal energy and pushes the gas via adiabatic expansion of the hot shocked wind bubble. 
\citet{Costa2014} have shown that the momentum-driven solution is much less efficient in cosmological simulations than in isolated halo simulations, whereas the energy-driven solution is proven to be efficient in driving outflows also in large-scale (and therefore low resolution) cosmological simulations.
Therefore we choose the energy-driven solution in our work.
However, a major drawback of the energy-driven solution is that the injected energy is instantly radiated away by strong cooling, which appears to be a numerical effect of the simulation \citep{Ceverino2009}.
While other mechanisms, such as cosmic rays or magnetic fields, with longer dissipative time scales are thought to sustain the pressure of the blast from this instant cooling {\citep{Cox2005,Salem2016,Hopkins2019}}, we choose to momentarily stop the cooling of the gas after the energy injection. 
This feature, called `delayed cooling', has been used in other works \citep{Stinson2006,Governato2010,Agertz2011}, and results in a temporary over-estimate of the temperature of the affected cells.
{
Those cells that are affected by the delayed cooling would, due to their artificially high temperature, cause a very strong line emission that is unrealistic.
At redshift 0.7 these cells make up only around 0.5\% of all cells in the {most massive} halo.
Therefore, in order to do an analysis on the line emission of gas around galaxies, we artificially remove these cells before post-processing the simulation snapshots.
}

\section{Flux emission prediction}

The objective of this work is to put together a realistic model for faint diffuse emission from the CGM in order to prepare observations of the CGM with upcoming instruments.
{The model is set up such that we can make mock observations for any emission line from the CGM, e.g. typical UV lines such as Ly$\alpha$ at 1216 \AA , OVI at 1032/1038 \AA\ or CIV at 1548/1551 \AA, optical lines such as H$\alpha$ at 6563 \AA\ or X-ray lines, such as OVIII at 19.0 \AA , CVI at 33.7 \AA\ or NeIX at 13.4 \AA .
While we create a general model for any emission line, later in the analysis we will focus mainly on the UV line Ly$\alpha$ for comparison with observations and preparation for FIREBall-2. 
For the predictions for HARMONI we will mainly consider (redshifted) UV and optical lines.
X-ray lines will be discussed in a subsequent publication.
}
It is beyond the scope of the present analysis to propose specific improvements of the complex emission mechanisms from the CGM.
Nevertheless, the high resolution reached on such large scale simulations brings innovative insight into CGM gas phase emission line physics.
We structure this section such that we first introduce the simple emission model applied to all emission lines for hydrogen and metals before we discuss some specifics of the complex Ly$\alpha$ emission.

\subsection{General emission model}

There are different mechanisms responsible for the expected extended CGM emission.
The first one, referred to as \textit{gravitational cooling}, is due to the collisional ionization of accreting gas, radiating away part of the energy acquired by compression and shock heating.

The second source is the photo-ionization by external UV sources, which causes the ionization of the gas and subsequent emission of photons via recombination processes (\textit{fluorescence}). 
Among these UV sources, there is the metagalactic UV background (UVB), which consists of the UV photons emitted from distant objects, such as stars or quasars \citep{Haardt2001,Haardt2012,Kollmeier2014}.
The computation of the UV backgrounds is a complex task, as many parameters come into play, such as the ionizing photon escape fraction as well as the dust content and opacity and the density distribution of absorbing gas \citep{Haardt2001,Kollmeier2014,Khaire2018}.
In addition to this metagalactic background, there are cases where the presence of a photo-ionizing bright source nearby, such as a quasar \citep{Cantalupo2005,Kollmeier2010,Cantalupo2014,Martin2014a} enhances the illumination of the gas locally, which then re-radiates through fluorescence.

In the following we investigate the relative contribution of these different sources to the total luminosity, which is an actively debated topic.
We discuss these different mechanisms in the context of our high-resolution simulation and how we take them into account in the post processing.

The exact determination of the contributions from these different sources requires on-the-fly calculation within the hydrodynamical simulation itself. 
This has been done for the UV ionizing continuum that impacts the ionization, the temperature and the dynamics of the gas, and consequently changes its emissivity \citep{Rosdahl2013}.
A good approximation of this on-the-fly UV ionizing photon transfer has been developed by \citet{Rosdahl2013}, namely the 'on-the-fly' self-shielding, used here in our high-resolution simulation.

\subsubsection{Emissivity tables}

Similarly to \citet{Bertone2010b} and \citet{Frank2012}, we generate emissivity tables for our lines of interest at the corresponding redshifts to attribute a luminosity to each gas cell. 
These tables account for the flux produced by the gravitational cooling of the gas, and the recombinations from the UVB photo-ionization.

We use the photo-ionization code CLOUDY, version 10.01\footnote{We are using this version of Cloudy as it includes the option to compile with double floats, which is not computed in the c13 Cloudy version. 
This feature is important in our case, as we are deriving the emissivity from low density regions. 
These regions can have emissivities below -32 dex.}, last described by \citet{Ferland1998}.
This code predicts the thermal, ionization, and chemical structure of a cloud illuminated in a variety of physical conditions.
{We want to note here that the cooling function in CLOUDY is more sophisticated than the simplistic model for cooling in RAMSES and the inconsistency between the two may introduce a bias in our predictions.
The temperature in the simulation will adjust so that the photon emission (especially Ly$\alpha$) accounts for the cooling required to roughly balance the total heating rate. 
Post-processing simulations with inconsistent cooling tables may result in luminosities greater than the heating rates that were present during the simulation.
It is however beyond the scope of this paper to investigate this uncertainty further.}

We consider a 1cm slab of optically thin gas at solar metallicity\footnote{The emissivity $\rm \epsilon$ scales linearly with metallicity in the first order. We tested this by running several models with varying metallicity, density and temperature and found a linear correlation between the metallicity and the emissivity on scales between 0.1 solar metallicties and 10 solar metallicities.}, with no molecules and using the element abundances in the solar photosphere as described by \citet{Grevesse2010}. 
In our model, we use the background derived by \citet{Haardt2001} (HM01 in the following) with contributions from both quasars and galaxies to be consistent with \citet{Frank2012}.

We derive the hydrogen density $\rm n_{H}=\frac{X_{\odot}}{m_{H}}\rho$, with $\rm X_{\odot}=0.7380$,
and the weighted temperature $\rm T/\mu=\frac{m_{H}}{k_{B}}\frac{P}{\rho}$ from the simulation using \citet{Grevesse2010} abundances (also used in the Cloudy models for consistency).

\begin{figure}
\begin{center}
\includegraphics[width=.5\textwidth]{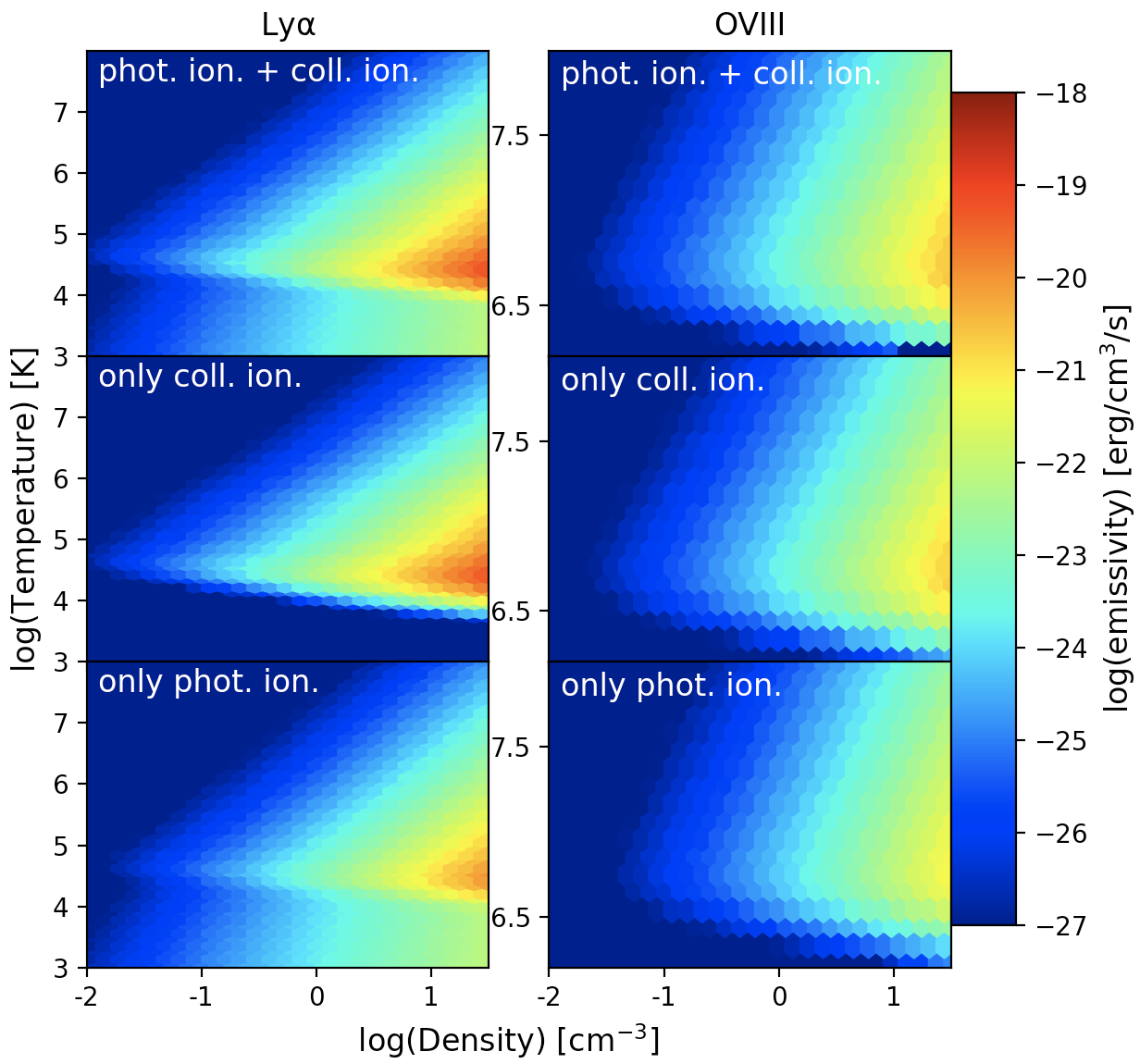}
\caption{{Examples of our CLOUDY emissivity diagrams} in units of erg/s/cm$^{3}$ in logarithmic scale for Ly$\rm \alpha$ ({left}) and OVIII ({right}) at a redshift of z=0.33. 
The temperature is given in $\rm T/\mu$.
{The density is given as total hydrogen density in at/cc.}
The {top} panels show the dual contribution of photo-ionization equilibrium (PIE) and collisional ionization equilibrium (CIE), while the {middle} panels show the sole contribution of the CIE, used for the self-shielded gas, {and the lower panels show the sole PIE contribution, for comparison}. 
} 
\label{im:cloudytables}
\end{center}
\end{figure}

\begin{figure*}
\centering
\includegraphics[width=.49\textwidth]{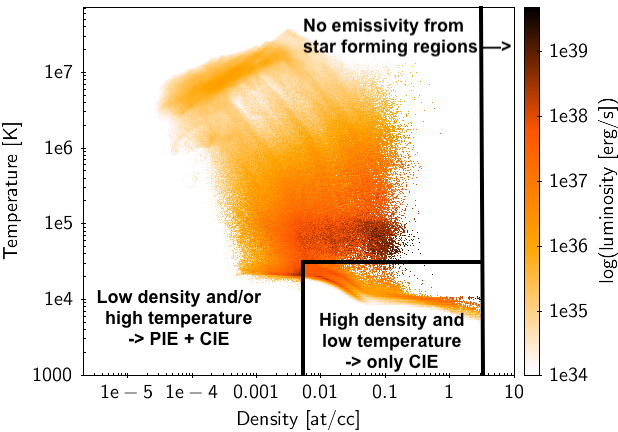}
\includegraphics[width=.49\textwidth]{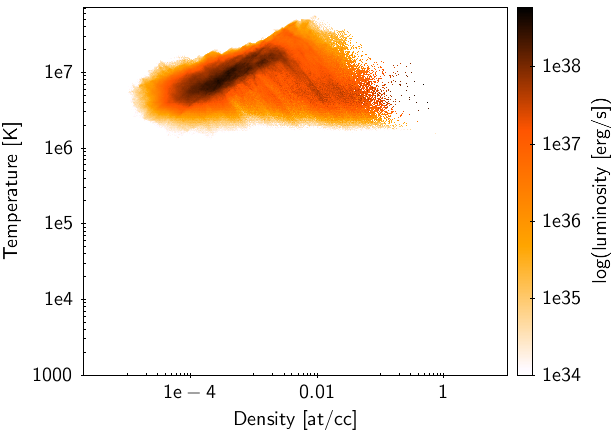}

\caption{{Lumonisities resulting from our emissivity predictions} applied to gas cells in the selected most massive halo at redshift 0.7. {The density is given as total hydrogen density in at/cc.}
The left panel shows the example of Ly$\alpha$. 
Star forming regions with a number density of $\rm n_{HI}$>3 at/cc are located within the ISM and are not relevant for our CGM emission study. 
They are therefore assumed to have no emission at all.
To account for the self shielding of gas clouds we apply the cut number 2 from \citet{Frank2012} to identify gas cells for which we only consider collisional ionization (CIE).
This leads to null emission from cold regions.
For all remaining gas cells, we consider emission from both photoionization (PIE) and collisional ionization (CIE).
The right panel shows the emissivity prediction of the same halo for OVIII emission.
}
\label{emissivityfromhalo}
\end{figure*}

The tabulation of $\rm log(T/\mu)$ is done in two steps. First, we generate emissivity tables, as well as electronic density tables in density-temperature (n,T). We use $\rm \log(n_{min})=-8$, $\rm \log(n_{max})=4$, $\rm d(log(n))=0.1$, $\rm \log(T_{min})=2$ and $\rm \log(T_{max})=8$, $\rm d(log(T))=0.1$. For each point (n,T) in the emissivity tables, we generate a new coordinate (log(n),$\rm log(T/\mu)$) using $\rm \mu=\frac{n_{H}A_{H}+n_{He}A_{He}}{n_{H}+n_{He}+n_{e}}$ where $A_{H}=1.0074$ is the mass number of hydrogen, $\rm A_{He}=4.002602$ the mass number of helium, and $\rm n_{H}$, $\rm n_{He}$ and $\rm n_{e}$ are the hydrogen, helium and electronic density respectively, tabulated along the emissivity tables. This gives a non-uniformly distributed (log(n),$\rm log(T/\mu)$) emissivity table which we then interpolate back on a regular grid using $\rm \log (T/\mu)_{min}=min(log(T/\mu))$ and $\rm \log(T_{max})=max(log(T/\mu))$ and $\rm d(log(T/\mu))=\frac{1}{3}d(log(T))$.
We then interpolate the different emissivity tables available to the corresponding expansion factor of the considered snapshot.

Figure \ref{im:cloudytables} shows {some examples of} the so created emissivity tables for 2 differents lines: $\rm Ly\alpha$ and OVIII.
The {top} panels show the joint contribution of Photo-ionization (PIE) and collisional ionization (CIE), while the {middle and lower} panels show the sole contribution of CIE {and PIE, respectively}.

In our model of all the lines, the main contribution to the total emission comes from collisional ionization.
Photoionization only plays a role at low temperatures and is negligible in all cases but $\rm Ly\alpha$.
Although this is the case in our models, other works find different solutions for the ionization contributions (e.g. \citealt{Cantalupo2017,Oppenheimer2018}), {suggesting that photoionization is the dominant source for emission from the CGM, rather than collisional ionization.}

There are regions in the density-temperature space that are unresolved by CLOUDY, because the occupation of certain states is unlikely and the calculation time intensive.
{Those are negligible and do not affect our results, as they would show only small emissivity if any.}
Also, the unresolved regions have generally low temperatures and high densities corresponding to cool ISM gas within the galaxies themselves, away from the CGM regions we are interested in.

\subsubsection{Post-processing self-shielding}
\label{secrtion:selfshielding}

In section 2.2 a calibrated empirical technique has been used to mimic the effect of self-shielding in the gas temperature and density.
Here we describe how we take the effect of self shielding into account in the post-processing.
Asserting the fraction of gas self-shielded from ionizing radiation is a rather delicate topic. 
In \citet{Frank2012}, the most optimistic self-shielding model uses \citet{Popping2009} results {$P/k > 258$ and $\tau_{rec} < \tau_{s}$, based on the equilibrium between sound speed of the gas and recombination time and an empirical constraint on the thermal pressure}.
{\citet{Frank2012} used this model due to the limiting resolution in their simulation. 
They present 9 different possible choices on the self-shielding thresholds, some of them only including few gas cells from their simulation.}
With our increased resolution {and higher threshold for star formation, we apply a different threshold for the self-shielding than \citet{Frank2012}.}

We adopt the model proposed by \citet{Furlanetto2004} that simply puts a condition on the temperature (the gas at $\rm T>10^{4.5}K$ is collisionally ionized and not self-shielded) and on the density. 
The density threshold of $\rm n_{HI}\sim 10^{-2}at/cm^{3}$ is in line with \citet{Rosdahl2012} (at z=3) and \citet{Faucher-Giguere2010} prescription from radiative transfer analysis.
\citet{Frank2012} have discussed different cuts for the self-shielding in the post-processing in more detail.
The conditions we choose for this work correspond to their cut number 2.
It considers the three regimes shown in the {left} panel of figure \ref{emissivityfromhalo}.
For the ISM region with  $\rm n_{HI} > 3$ at/cc, we consider zero emission.
The self-shielded gas with only collisional ionization (CIE) is defined for $\rm n_{HI} > 5.1 \times 10^{-3}$ at/cc and $\rm T< 10^{4.5} $ K.
The rest of the gas emits through both CIE and photoionization (PIE).

\subsubsection{Non-thermal feedback}

For many star forming galaxy halos in our high-resolution simulation, we find a small percentage ($\rm <1\%$) of gas cells with delayed cooling (from the non thermal feedback).
These cells reach temperatures of $\rm 10^{5-6}K$, with densities consistent with ISM gas cell ($\rm n_{H}>0.1~at/cm^{3}$). 
{
They have emissivities of several orders of magnitudes above the value we would expect if their temperature had not been artificially increased, and have therefore also a much higher luminosity than what would be realistic. 
}
{We have tested the sensitivity of the total halo luminosity against the amount of cells we exclude of the simulations. If we remove more cells than just the ones in delayed cooling, the total luminosity of the halo remains unchanged. If we, however, leave some of the delayed cooling cells in the halo, the luminosity rapidly increases by an order of magnitude. Therefore we conclude that the exclusion of the delayed cooling cells is a conservative approach.}

We choose not to consider these particular cells in the total luminosity budget, as they would in reality not reach these high temperatures but higher pressures, and for shorter time scales, more as a flash. This consideration brings no particular bias in the total luminosity budget (we remain conservative by not taking them into account), as we checked that these cells, originally associated with ISM gas, should not contribute predominantly to the CGM emission.

\subsection{Special treatment of Ly$\alpha$}

For Ly$\alpha$ we use the CIE and PIE emission tables just as for the metals.
While there is some debate about the dominant source spatially extended Ly$\alpha$ emission (e.g. \citealt{Cantalupo2017}), collisional ionization is thought to be a main contributor (about 50\% of this cooling radiation emerges as Ly$\alpha$ photons) as the photons thus created would be emitted in the dust-poor outskirts of the disc, hence only a small part of these photons is affected by the subsequent immediate dust absorption \citep{Fardal2001,Dijkstra2009,Faucher-Giguere2010}.

An additional contribution comes from the production of ionizing photons from star formation or a quasar inside the halo. 
Indeed, $\rm h\nu > 1 Ryd$ photons can ionize the ISM gas, producing \textit{photon scattering} out of the star forming regions, although this contribution is mainly Ly$\alpha$ and negligible for metal lines.
We take this into account by creating a simple model for the galaxy Ly$\alpha$ emission based on the SFR.

To reproduce the diffuse Ly$\alpha$ line emission, on-the-fly radiative transfer is not essential as the post-processing of the transfer of the resonant Lya photons gives reliable estimates of the total flux emitted \citep{Verhamme2006,Kollmeier2010,Faucher-Giguere2010,Trebitsch2016}.

{
\subsubsection{Induced Processes}

By default, CLOUDY takes into account induced processes: induced recombination and its cooling, stimulated two-photon emission and absorption \citep{Bottorff2006}, continuum fluorescent excitation, and stimulated emission of all lines ({\it Hazy - a brief introduction to CLOUDY C10 - 1. Introduction and commands}\footnote{https://www.nublado.org/}, p.237). 
The {\it no induced option} turns all these processes off and has also been used in \cite{Frank2012} as well as in \cite{Furlanetto2004,Furlanetto2005}.
For a full explanation for this choice we refer the reader to the respective works, but we highlight two of the main reasons:

$\bullet$ The absorption and immediate re-emission of isotropically distributed Ly-$\alpha$ photons do not contribute to the net luminosity and any excess luminosity due to these processes as calculated in CLOUDY are therefore subtracted from the total emissivity of a gas cell.

$\bullet$ CLOUDY allows for an unphysical conversion of an absorbed Ly-$\beta$ photon ($1s \rightarrow 3p$) to an emitted Ly-$\alpha$ photon ($2p \rightarrow 1s$) because it assumes mixed orbital angular momentum states for $n \geq 3$. This assumption is motivated for extremely high gas densities ($ n_{H} \gg 10^{8} cm^{-3}$, \citealp{PengellySeaton1964}) and therefore not valid in our case.

However, here we explore the actual difference between using induced processes and using the 'no induced processes' option.
We note that one of the contributions to the induced processes is the {\it photon pumping} or scattering of continuum photons to the Ly-$\alpha$ line. 
This contribution is heavily dependent on the geometry of the gas cloud and therefore likely unconstrained with our simplistic CLOUDY setup. 
It also depends on nearby ionizing continuum objects, such as young stars or AGN (see e.g. \citealt{Cantalupo2017}).
We rather model the effect from the nearby stars separately (see the next section). 

If we run CLOUDY with induced processes and apply it to a simulated halo, the total luminosity in the halo is a factor 2-6 higher than when applying a CLOUDY model with 'no induced processes'.
While we choose to use the option 'no induced processes' in our analysis for the above mentioned reasons and to stay consistent with previous works, this choice will result in a conservative estimate in our predicted halo fluxes.

\subsubsection{Ly$\alpha$ scattering from nearby stellar continuum}
}
In addition to {the photons from the UV background,} accounted for in \citet{Haardt2001} (HM01), ionizing photons ($\rm h\nu  \geq 1 Ryd$) emitted by nearby young stars, in particular these belonging to the halo in consideration, {can} contribute substantially to the total Ly-$\alpha$ emission of star forming galaxies. 
{The strength of this contribution depends strongly on the interstellar dust and gas geometry and kinematics \citep{Kunth2003,Verhamme2012}, as those determine how many Ly-$\alpha$ photons escape from the star forming regions.}
In a first, conservative approximation, we assume that all the ionizing photons from stars are absorbed by dust or photoionizing neutral gas in the ISM.
Since the dust attenuation is poorly constrained at low redshift, we will consider a simplistic model for the emitted Ly$\rm \alpha$ photons. 
We start with the prescription from \citet{Furlanetto2005} to estimate the intrinsic Ly$\alpha$ luminosity from ionizing photons in the absence of dust:
$ \rm L^{stars}_{Ly\alpha} [erg/s] = 10^{42} SFR [M_{\odot} yr^{−1}]$. We compute the SFR of each halo from the mass of young stars, using the `continuous star formation' approximation \citep{Kennicutt1998}:
\begin{equation}
\rm SFR [M_{\odot}/yr] = \dfrac{M_{stars < 10^8 yrs} [M_{\odot}]}{10^8 [yr]} 
\label{eq:sfr}
\end{equation}
 Using COS data of low redshift ($\rm  z \sim 0.03$) star forming galaxies, \citet{Wofford2013} measured a Ly$\alpha$ escape fraction ranging from 1 to $\rm 10\%$. 
 They estimate that this fraction is sensitive to the presence of dust and to the HI column density, the Ly$\alpha$ photons escaping more easily from holes of low HI and dust column densities, resulting in a large scatter.
 Winds can also have a strong effect and can help Ly$\alpha$ photons to escape \citep{Dijkstra2013} but we do not consider winds in our model.
As we do not have any model for either the dust or radiative transfer, we will stay conservative in our assumptions.
\citet{Hayes2011} find  Ly$\alpha$ escape fractions between $\rm 0.1\%$ and $\rm 1\%$ for low-z galaxies.
Therefore, we will consider two extreme cases for the Ly$\alpha$ luminosity from the stellar contribution at low redshift: one with a Ly$\alpha$ escape fraction of $\rm 1\%$, and another with a Ly$\alpha$ escape fraction of $\rm 0.1\%$. 
At higher redshift ($z$>1), we adopt a Ly$\alpha$ escape fraction of $\rm 10\%$ which is within the prediction of \citet{Hayes2011}.
{
We are aware that different works predict very different escape fractions for Ly$\alpha$ (e.g. \citealt{Wofford2013,Naidu2017}, but we choose to follow the trend observed in \citet{Hayes2011}.
}
Predicting the spatial and spectral profiles of such emission would require the full calculation from radiative transfer techniques \citep{Verhamme2006,Verhamme2012,Rosdahl2013,Lake2015}, which is beyond the scope of this work. 
However, as our goal is to study the detectability of such emission with upcoming instruments, we chose to make the simple assumption that all the Ly$\alpha$ photons only go through one aborption/re-emission process before leaving the cloud. Also, we assume that all of the Ly$\alpha$ photons are emitted from the centre of the galaxy. 
We then weigh the profile proportionally to the total hydrogen density of the gas cell and by its inverse squared distance to the centre. This gives us, for each cell j the luminosity $\rm L_{j}^{\star}$:
\begin{equation}
\rm L_{j}^{\star} [ergs/s] = f_{esc} (Ly\alpha) \dfrac{\frac{n_{j,H}}{R_j^2}}{\int_{R_{vir}} \frac{n_H}{R^2}} 10^{42} SFR_j [M_{\odot}/yr]
\end{equation}

\subsubsection{Total Ly$\alpha$ Luminosity}

As described in previous sections, we consider collisional ionization (gravitational cooling) and photo-ionization from UV background photons.

We consider that the total luminosity for the Ly$\alpha$ line is the sum of these two quantities:
\begin{equation}
\rm L^{total}_{Ly\alpha} = L^{grav. cooling + UVB fluo.}_{Ly\alpha} + L^{stars}_{Ly\alpha} 
\end{equation}
This consideration is not completely realistic, as we should strictly take into account the ionizing flux from the young stars combined with the UVB used in the Cloudy model and during the simulation computation to reproduce the density/temperature state of the gas in these conditions. 
The RAMSES simulation used in this work only reproduces the gravitational effects and the heating from the UVB. Regarding the purpose of the present work, this assumption should be accurate enough to give valuable insights on the level of radiation from the CGM.

\subsection{UV continuum}

To properly reproduce mock observations in the UV, we now need to model
the UV continuum of each halo. We first compute the SFR of each halo from
the mass of young stars (see eq. \ref{eq:sfr}), from which we infer the flux derived by \citet{Kennicutt1998} , $\rm L (\lambda) [erg/s/\AA] = \dfrac{SFR [M_{\odot}/yr]}{1.4} \dfrac{c}{\lambda^2} 10^{28}$. To derive the spatial extent of
this continuum, we assume that the UV continuum is mainly produced by these young stars, so we use the stars whose age is less than $\rm 10^{8}$ years to derive a `stellar density field' that we scale with $\rm L_{\lambda,1500 \text{\AA}}(\lambda)$:
\begin{equation}
\rm L_{\lambda,1500 \text{\AA}}(\lambda) [erg/s/\AA] = 10^{−0.4 k^{\prime} E(B-V)} \dfrac{SFR [M_{\odot}/yr]}{1.4} \frac{c}{\lambda^2} 10^{28}
\end{equation}
To account for the dust attenuation of these continuum photons, we use the model from \citet{Zahid2012} to get the color excess $\rm E(B-V)$ from the sum of the stellar particles (not just the young stars) using their mass $\rm M_{\star}$ and metallicity Z:
\begin{equation}
\rm E(B-V)=0.44(p_0 +p_{1}Z^{p_{2}})M^{p_{3}} 
\end{equation}
where $\rm Z = 10^{(12+log(O/H)−8)}, M = M_{\star}/10^{10}, p_0 = 0.12 \pm 0.01, p_1 = 0.041 \pm 0.006, p_2 = 0.77 \pm 0.06, and \  p_3 = 0.240 \pm 0.002$. 
{This model from \citet{Zahid2012} has been fitted to SDSS data to determine those parameters. The fit rms is $\sim$ 0.11 dex. Particularly, the high stellar mass - high metallicity, and therefore high colour excess, objects from the SDSS observations are slightly underpredicted in their work, which may affect our mock observations such that the color excess for the {most massive} halos is slightly underestimated.}
We consider the stellar mass weighted metallicity to derive Z from the simulation. We derive the extinction $\rm k^{\prime}(\lambda)$ following \citet{Calzetti2000}:
\begin{equation}
\rm k^{\prime} (\lambda) = 2.659(- 2.156 + 1.509/\lambda - 0.198/\lambda^2 + 0.011/\lambda^3) + R^{\prime}_{V}
\end{equation}
for $\rm 0.12\mu m \leq \lambda \leq 0.63 \mu m$.
We choose to use $\rm R_{V^{\prime}} = 3.1$ to account for a dusty environment such as the Galactic diffuse ISM. The attenuation of the continuum then scales as $\rm 10^{−0.4 k^{\prime} E(B-V)}$. We make the assumption that the attenuation is homogeneous within the ISM {across the projected image of its UV continuum}.

\section{Results}

\begin{figure}
\centering
\includegraphics[width=0.49\textwidth]{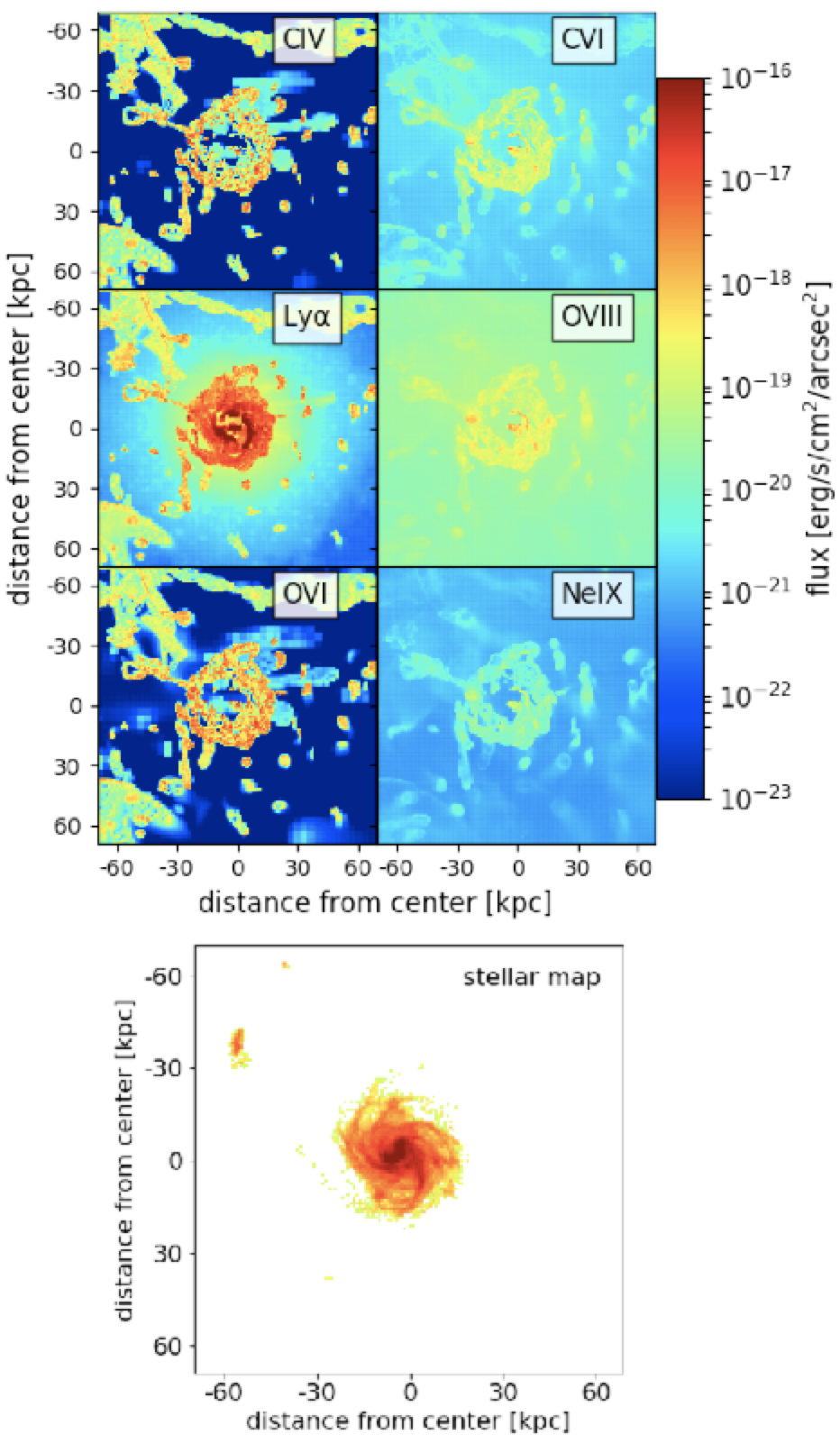}
\caption{{Upper} Panel: Simulated flux emission predictions of various lines at redshift 0.3. Pixel scale is $\rm 0.2^{\prime\prime}$ corresponding to $\rm 0.9 kpc$ at the given redshift.
We find Ly$\alpha$ to be the brightest emission line among the lines emitting at UV wavelengths (left panels) and OVIII to be the brightest among the X-ray lines (right panels).
We show all lines with the same color scale to illustrate the distribution of the different gas phases in the CGM. 
The emission from colder gas (lower ionization states) comes from clumpy structures in the CGM with only little flux from the areas in between. 
The hot component, traces the same structures but with a much higher flux level also in between the individual dense gas clumps, such that a more homogeneous distribution of flux over the entire halo becomes apparent. 
{See also Figure \ref{im:xrayvsuv} in the appendix for a comparison between Ly$\alpha$ and OVIII emission on larger scales.}
{Lower} Panel: Stellar map of the {most massive} halo at redshift 0.3. 
}
\label{im:mainhaloions}
\end{figure}

In this section we present the results of our simulations and compare them to observations in order to estimate how realistic our predictions are.
We note that the main aspect of our work is the emission prediction of galaxy halos rather than the overall properties of the cosmological simulation itself.

\begin{table*}
\centering
\caption{{Most massive} halo luminosity and integrated flux {predictions from our simulations} for different ions and redshifts {for a given Ly$\alpha$ escape fraction}. 
}
\label{tab:haloluminosities}
\begin{tabular}{cccccccccc}
\hline\hline
z & $\rm f_{esc}$ & $\rm L_{Ly\alpha}$ & $\rm log(f_{Ly\alpha})$ & $\rm L_{CIV}$ & $\rm log(f_{CIV})$ & $\rm L_{OVI}$ & $\rm log(f_{OVI})$ & $\rm L_{OVIII}$ & $\rm log(f_{OVIII})$  \\ 
 &  & [$\rm 10^{42} erg/s$] & [$\rm erg/s/cm^{2}$] & [$\rm 10^{42} erg/s$] & [$\rm erg/s/cm^{2}$] & [$\rm 10^{42} erg/s$] & [$\rm erg/s/cm^{2}$] & [$\rm 10^{42} erg/s$]  & [$\rm erg/s/cm^{2}$] \\\hline
 4.0 & $\rm 10\%$ & $\rm 10.0$ & $\rm  -16.2$  & $\rm 0.71$ & $\rm -17.4$  & $\rm 0.92 $  & $\rm -17.2$ & $\rm 0.05$  & $\rm -18.5$\\
 2.3 & $\rm10\%$ & $\rm 35.9$ & $\rm  -15.1$  & $\rm 1.17$ & $\rm -16.6$  & $\rm 1.44 $  & $\rm -16.5$ & $\rm 0.35$  & $\rm -17.1$\\
 1.0 & $\rm 1\%$ & $\rm 7.93$ & $\rm -14.8$ & $\rm 1.56$ & $\rm -15.5$  & $\rm 1.49$  & $\rm -15.6$ & $\rm 0.64$  & $\rm -15.9$\\
 0.67 & $\rm 1\%$ & $\rm 3.73$ & $\rm -14.7$ & $\rm 1.16$ & $\rm -15.2$  & $\rm 1.16$  & $\rm -15.5$ & $\rm 0.47$  & $\rm -15.6$\\
 0.25 & $\rm 0.1\%$ & $\rm 2.14 $ & $\rm -13.9$& $\rm 1.18$ & $\rm -14.2$& $\rm 1.17$ & $\rm -14.2$& $\rm 0.51$  & $\rm -14.6$\\
\hline
\end{tabular}
\end{table*}

\subsection{Different ions in the {most massive} halo at z=0.3}

Once we apply the emission prediction model to galaxy halos from the simulation, we can calculate the luminosity and the flux in each cell for a given ion at a given wavelength.
We use these calculated fluxes to create data cubes of these halos that represent mock observations with two spatial axes and one spectral axis.
First of all we look at the qualitative difference of emission from different ions from the {most massive} halo at a given redshift.
We consider the {most massive} halo at redshift 0.3.
Fig. \ref{im:mainhaloions} shows the outcome of the emission prediction for the {most massive} halo at different wavelengths, corresponding to different ions.
We see that the gas emitting in Ly$\alpha$, CIV and OVI lines (left side panels) is much more concentrated in clumps than the hot gas emitting CVI, OVIII and NeIX lines on the right side panels which seem more homogeneously distributed around the central galaxy.
We also find that Ly$\alpha$ is the overall brightest line amongst the UV lines and OVIII the most promising X-ray line from high temperature gas.

\subsection{Comparison of the {most massive} halo luminosity to the low-resolution simulation near z $\sim$ 0.67}

We note that the maximum Ly$\alpha$ luminosity {at low redshifts} (see Tab. \ref{tab:haloluminosities}) does not exceed a few $\rm 10^{42} erg/s$.  
{Yet,} our high-resolution simulation is based on one of the brightest {and most massive} halos from the analysis performed by \citet{Frank2012}. In their analysis {for Ly-$\alpha$ at z$\sim$0.67}, they predict Ly$\alpha$ luminosities to go up to $\rm 10^{44}\ erg/s$, without even accounting for the SFR induced Ly$\alpha$ luminosity. This two dex difference finds its origin in the recipe used for the high-resolution simulation. Indeed, while we used 'on-the-fly' self-shielding in our simulation, preventing $\rm n_H > 10^{-2}\ at/cc$ gas cell to be heated by the metagalactic UVB, the gas cells in \citet{Frank2012} show larger temperature in the cooling 'gutter' of Ly$\alpha$ emission (see Fig. \ref{im:densitytempdiagramzoomedhalo}). 
{Specifically, we find the temperature in our cooling gutter at n=0.03 at/cc around $10^{4}$ K, while the temperature in the cooling gutter in the low resolution simulation at the same density is around $2 \times 10^{4}$ K.}
This low increase of the equilibrium temperature has dramatic effects on the effective emission rate, as there is a steep evolution of the Ly$\alpha$ cooling emissivity with temperatures of a few $10^4$ K ($\rm 10^2 \epsilon(10^4\ K) \approx \epsilon(2 \times 10^4\ K)$), see Figure 6 in \citet{Rosdahl2012}. We therefore argue that the emission level predicted in our model is more realistic, despite being less optimistic, than those derived in the original study as the crucial question of cooling temperature has been optimised since the last implementation of the code.

\subsection{Validation with low redshift observations}

\begin{figure}
\begin{center}
\includegraphics[width=.5\textwidth]{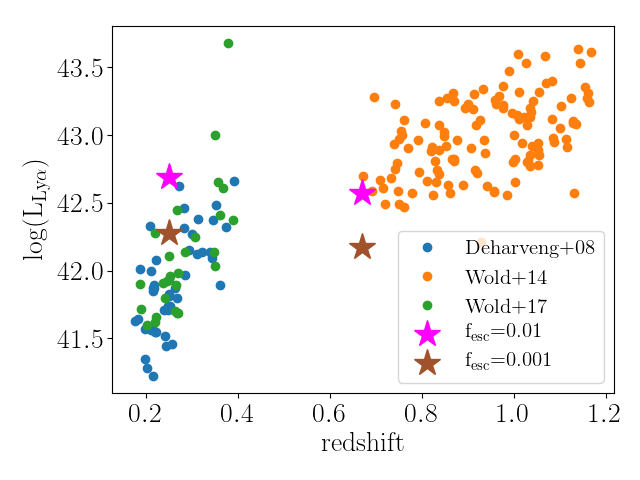}
\caption{Comparison of the predicted Ly$\alpha$ Luminosity at low redshifts with GALEX observations for the {most massive} halo as described in table \ref{tab:haloprops} and different Ly$\alpha$ escape fractions.
We plot the most massive halo and so {one of} the most luminous. We adjust the escape fraction such that this most massive halo fits well into the observations, which are also the most luminous ones.
} 
\label{im:galexcomparison}
\end{center}
\end{figure}

We want to see how well our simulations compare to actual observations at low redshift.
This is important not only to verify how realistic the simulations are but also to see how well they are suited for the preparation of future observations with upcoming UV instruments.
For this exercise we compare the simulated flux in the {most massive} halos at redshift 0.3 and 0.67 with GALEX observations of Ly$\rm \alpha$ emitting galaxies \citep{Deharveng2008,Wold2014,Wold2017}.
Since the Ly$\rm \alpha$ escape fraction at low redshifts can vary between 0.1\% and 1\% we consider both of these escape fractions \citep{Hayes2011}.
Fig. \ref{im:galexcomparison} shows this comparison.
For redshift z > 0.5 we are generally underestimating the observed flux, although our simulations are still in agreement with observations for $\rm f_{esc}=1\%$.
For the lowest redshifts (z<0.5) we are slightly overpredicting the flux and the flux for $\rm f_{esc}=0.1\%$ is only marginally in agreement with the observations.
{The {most massive} halo in our simulation, {chosen due to its high resolution (~380pc)}, is one of the most luminous due to its mass and size.}
The GALEX observations are, due to the instrument's detection limit, the most UV luminous galaxies at low redshifts, while there probably are many more less luminous galaxies at these redshifts.
In this context we conclude that our simulations are in agreement with {the observed Ly-$\alpha$ luminosities at low redshifts}.
Given the results of this comparison we {choose} a Ly$\rm \alpha$ escape fraction of $\rm f_{esc}=0.1\%$ for z < 0.5 and $\rm f_{esc}=1\%$ for 0.5 < z $\leq$ 1.0.

\subsection{Comparison to high redshift observations}

We consider recent observations of Ly$\alpha$ emission from high redshift Ly$\alpha$ halos from the Subaru telescope (z=2.3, \citealt{Momose2014}) and VLT/MUSE (z=4.0, \citealt{Wisotzki2016}) to validate our model. 
{Although the model described in the previous sections is originally set up such that it represents low-redshift ($z \leq 1$) objects, we use the same prescription for higher redshifts.
We test it at two example redshifts: z=2.3 and z=4.0.
Using this model for higher redshifts}
does not bring major changes in the post-processing self-shielding treatment, as the density cut {has been calibrated from} high redshift simulated galaxies(\citealt{Rosdahl2012}; {see also \citealt{Katz1996,Schaye2001}}) and the temperature cut is purely empirical. 
The dust attenuation calculation for the continuum is also considered redshift independent as the properties of dust grains should not evolve much. However, the Ly$\rm \alpha$ escape fraction $\rm f^{Ly\alpha}_{esc}$ is redshift dependent, and can be of the order of 10\% at redshift 4 \citep{Hayes2011}.

We note here that the comparison is a simplified case study of single objects.
This is due to scarcity of observations of galaxies with the properties that we require to do a meaningful comparison.
However this single object case study is appropriate to estimate the reliability of our simulations.

\subsubsection{Surface Brightness Profiles at z=2.2}

\begin{figure*}
\begin{center}
\includegraphics[width=.49\textwidth]{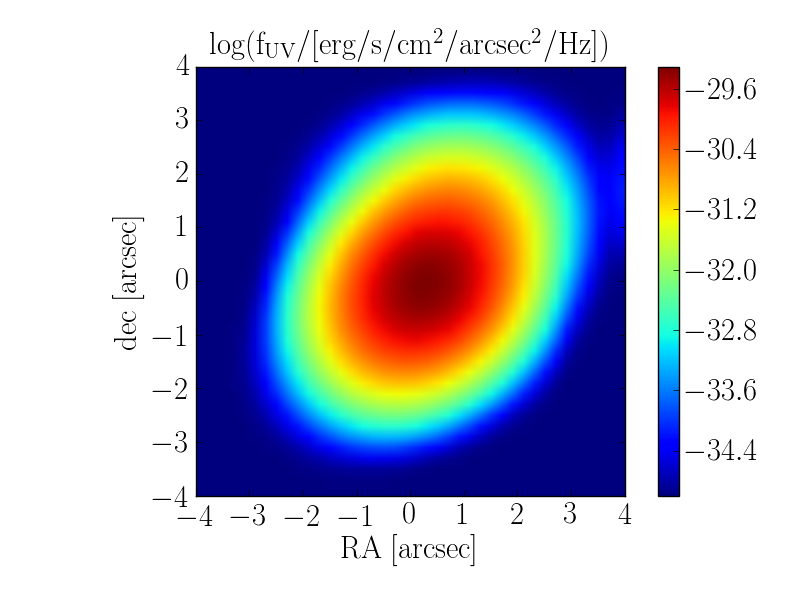}
\includegraphics[width=.49\textwidth]{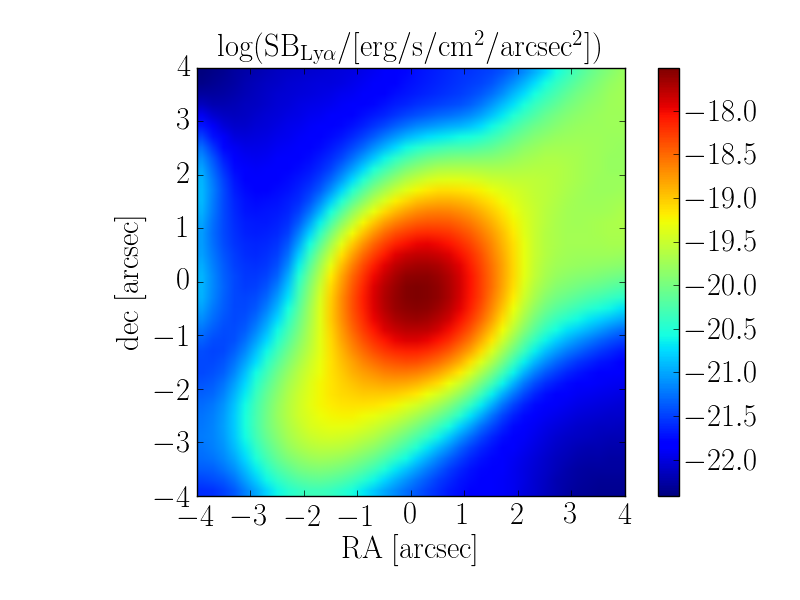}
\newline
\includegraphics[width=.49\textwidth]{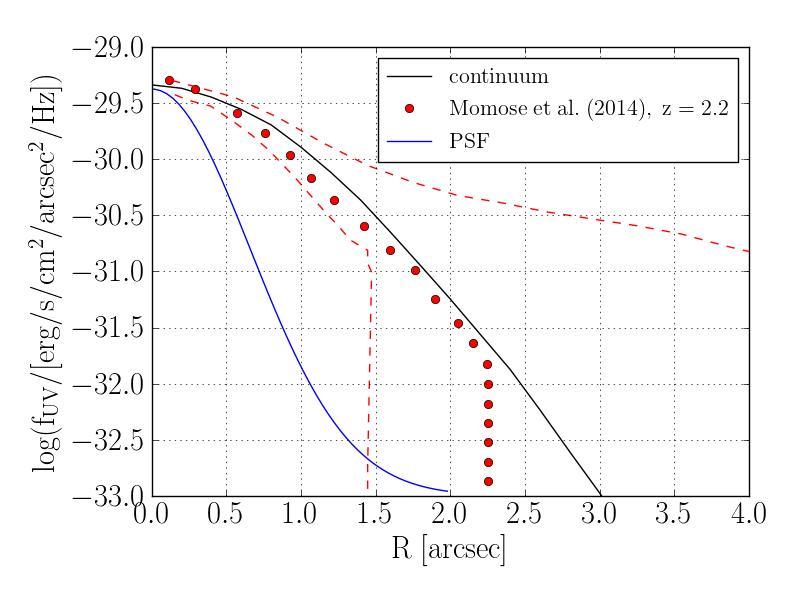}
\includegraphics[width=.49\textwidth]{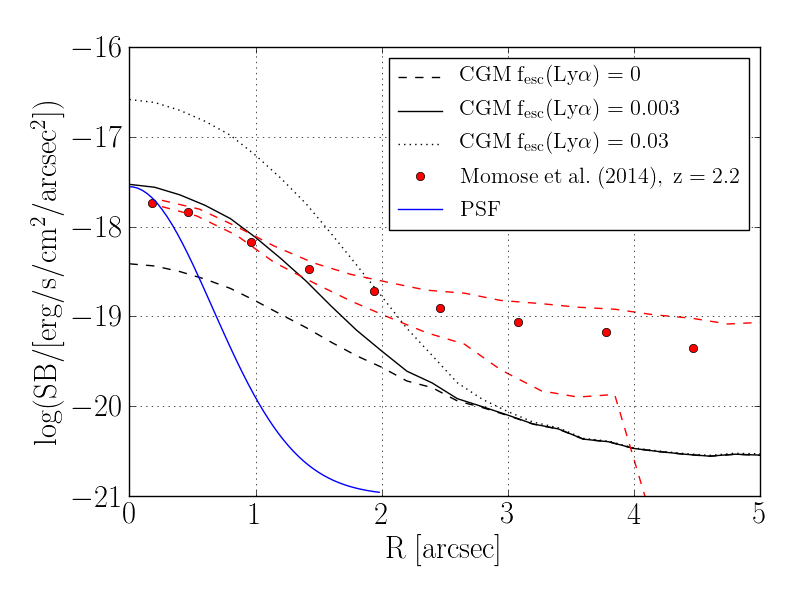}
\caption{Top left panel: surface brightness map of the continnum of the halo selected for the comparison with the stack of 3556 LAEs from \citet{Momose2014} at z=2.2. Bottom left panel: SB radial profile of the continuum for the selected halo and the stack. Top right panel: surface brightness map of the Ly$\alpha$ line emission, using $\rm f_{esc}(Ly\alpha) = 0.003$, for the halo selected in our study. Bottom right panel: SB radial profile of the Ly$\alpha$ line emission for the selected halo with Ly$\alpha$ escape fractions of $\rm \{0, 0.3, 3\}\%$ and for the stack.
The observed halo has a brighter CGM from R = 1.5 arcsec onwards, although the observed data points also become noisier in those outskirts.
{The red dashed lines give the uncertainty envelope on the observations within 1$\sigma$.
At the considered redshift of 2.2, 1 arcsec corresponds to 8.4 kpc.
The blue curve shows the shape of the PSF (FWHM=1.32 arcsec) with which we convolved our simulated data.}}
\label{im:sbz2}
\end{center}
\end{figure*}

The surface brightness (SB) profile at z=2.2 performed by \citet{Momose2014} consists in the stacking of 3556 LAEs, the comparison to one of our objects is therefore only illustrative. 
We identify a halo with $\rm M_{\star} = 4.8 \times 10^{10} M_{\odot}$ and $\rm SFR = 91.9 M_{\odot}/yr$ in the simulation, which reproduces a continuum level similar to that of the stack.
The top left panel of Fig. \ref{im:sbz2} shows the SB map of the continuum of the selected object, while the bottom left panel shows its SB radial profile with a comparison the the stack. As in the analysis by \citet{Momose2014}, we convolved the image with a PSF of 1.32 arcsec FWHM to reproduce the largest seeing size of the stacked images.
The top right panels shows the SB map of the selected object Ly$\alpha$ line, with the same convolution than the continuum. The bottom right panel shows the SB radial profile for the simulated object using Ly$\alpha$ escape fractions of $\rm \{0, 0.3, 3\}\%$ and that of the stack. 
For R < 1.5 arcsec, we are able to reproduce the Ly$\alpha$ line level with a Ly$\alpha$ escape fraction of $\rm 0.3\%$ (which is ten times below the prescription from \citet{Hayes2011} at this redshift), which corresponds to a Ly$\alpha$ luminosity of $L_{Ly\alpha} = 3.46 \times 10^{41} erg/s$.

We associate this under-estimation of the Ly$\alpha$ escape fraction at small radii with the uncertainties in the simulation and the uncertainties in the estimation of the Ly$\alpha$ escape fraction itself \citep{Hayes2011}.
Also the effect of stacking in \citet{Momose2014} as well as the SFR of the chosen halo (see \citealt{Matthee2016}) can play a role in this discrepancy.
We conclude overall that our simulated profile is comparable to observations.

At higher radii {(r$\gtrsim$15 kpc)} there seems to be an offset, which could mean an underprediction of the CGM flux in our simulation.
This would not be surprising, given that we do not use AGN feedback which is expected to drive more matter outside of the galaxy into the CGM.
{Also the lack of cosmic rays in our simulation may have a role in this discrepancy.
\cite{Hopkins2019} have recently shown that cosmic rays can have a significant impact on the CGM, especially at radii of r$\gtrsim$200 kpc as they keep cool gas from raining onto the galaxy.}
Yet, the observations at these larger radii are relatively uncertain and are not sufficient to draw a strong conclusion at this point.
For our purposes, the CGM flux is reproduced well enough in our simulations.

\subsubsection{Surface Brightness Profiles at z=4}

\begin{figure*}
\begin{center}
\includegraphics[width=.49\textwidth]{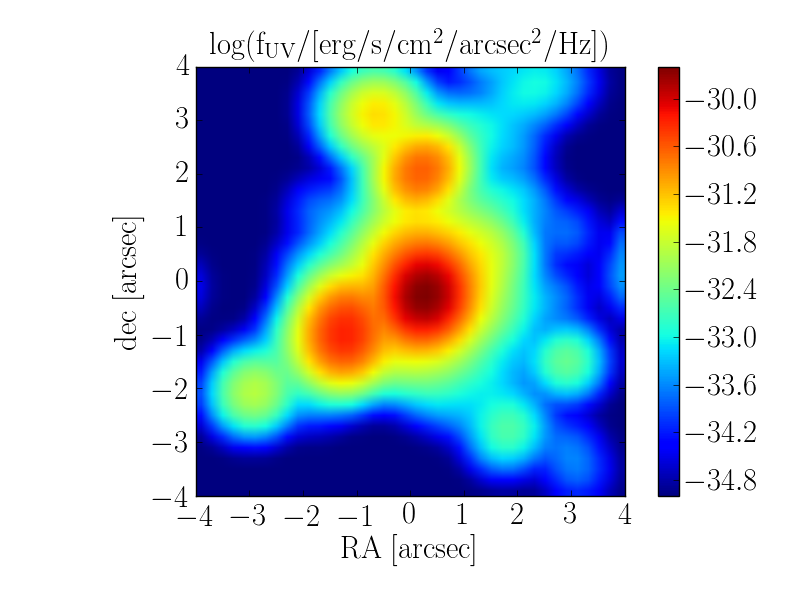}
\includegraphics[width=.49\textwidth]{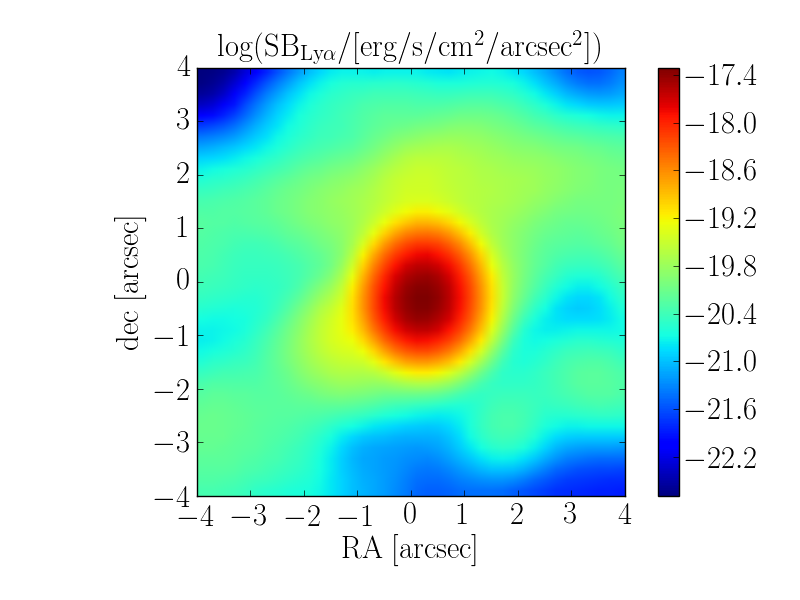}
\newline
\includegraphics[width=.49\textwidth]{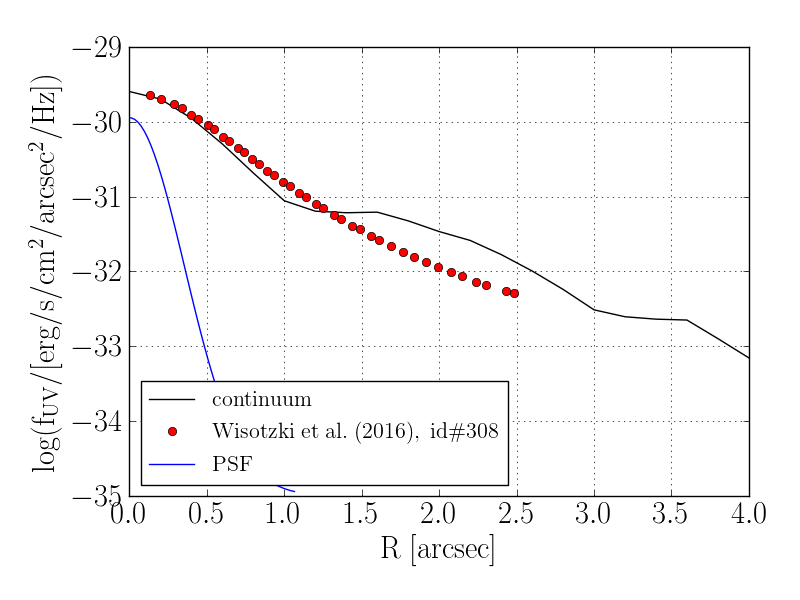}
\includegraphics[width=.49\textwidth]{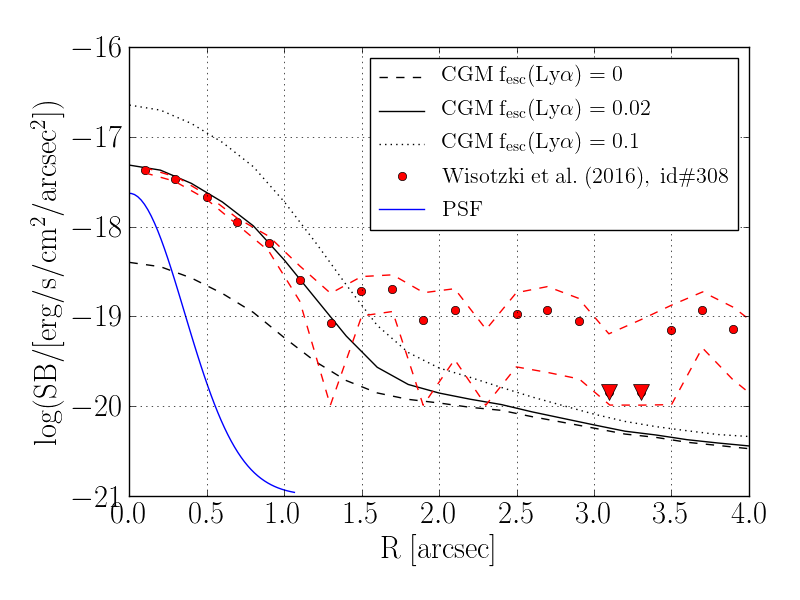}
\caption{Top left panel: surface brightness map of the continuum of the halo selected for the comparison with object \#308 from \citet{Wisotzki2016} at $\rm z=4.018$. Bottom left panel: SB radial profile of the continuum for the selected halo and object \#308. Top right panel: surface brightness map of the Ly$\alpha$ line emission, using $\rm f_{esc}(Ly\alpha) = 2\%$ for the selected halo. Bottom right panel: SB radial profiles of the Ly$\alpha$ line emission for the selected halo with Ly$\alpha$ escape fractions of $\rm \{0, 2, 10\}\%$ and for object \#308. The detection limit of the observed data is around log(SB)$\sim$-19.
{The red dashed lines give the uncertainty envelope on the observations within 1$\sigma$ and are statistically consistent with our predictions given the assumptions in the model.
At the considered redshift of 4.0, 1 arcsec corresponds to 7.1 kpc.
The blue curve shows the shape of the PSF (FWHM=0.66 arcsec) with which we convolved our simulated data.}}
\label{im:sbz4}
\end{center}
\end{figure*}

We chose the object \#308 from \citet{Wisotzki2016} for the comparison, as it lies at a redshift matching our high-resolution simulation set. We select a halo from the simulation based on the continuum SB profile which reproduces the continuum level of the object. The selected halo has a stellar mass $\rm M_{\star} = 7.0 \times 10^9 M_{\odot}$ and a SFR of $\rm 50.9 M_{\odot}/yr$, which is slightly above the prescription from \citet{Wisotzki2016} ($M_{\star} = 10^{8-9} M_{\odot} $ and $\rm SFR = 0.3 - 16 M_{\odot}/yr$).
The top left panel of Fig. \ref{im:sbz4} shows the SB map of the continnum of the selected halo. We convolved the image with a 0.66 arcsec FWHM PSF to account for the seeing and with a 0.71 arcsec FWHM PSF to reproduce the instrument's resolution \citep{Bacon2014}. The bottom left panel shows the SB radial profile for the selected halo and object \#308.
The top right panel shows the SB map of the Ly$\alpha$ line for the selected halo, with the same convolutions as the continuum. The bottom right panel shows the SB radial profile for the simulated object using Ly$\alpha$ escape fractions of $\rm \{0, 2, 10\}\%$ and that of object \#308. 
We recover a similar Ly$\alpha$ luminosity than the one measured by \citet{Wisotzki2016} for object \#308 ($\rm L_{Ly\alpha} = 1.6 \times 10^{42} erg/s$) with $\rm f_{esc}(Ly\alpha) = 2\%$: $\rm L_{Ly\alpha} = 2.0 \times 10^{42} erg/s$. 
This Ly$\alpha$ escape fraction is a few times lower than what \citet{Hayes2011} find at this redshift but we accept this difference as both the profiles as well as the Ly$\alpha$ escape fraction determination have some uncertainties associated.
Again, at larger radii (R > 1.5 arcsec), the simulated profile is lower than the observed one.
We note here that while the deepest MUSE observations can reach a detection limit of 2.8 $\times$ $\rm 10^{-20} erg/s/cm^{2}/arcsec^{2}/$\AA\ \citep{Leclercq2017}, the {specific} observations we compare with reach their detection limit at $\rm 10^{-19} erg/s/cm^{2}/arcsec^{2}$ and are therefore not easily comparable to the simulations at large radii.

\section{low redshift ($z\sim 0.7$) UV observations with FIREBall-2}

FIREBall-2 (Faint Intergalactic Redshifted Emission Balloon-2; PI: Chris Martin; \citealt{milliard10,Picouet2018}) is a balloon-borne experiment aiming at observing the faint diffuse UV emission from the CGM of intermediate redshift (0.3-1.0) galaxies. 
It consists in a UV Multi Object slit Spectrograph (MOS) with a resolution of $\rm R\sim2000$, and a FWHM of $\sim$ 6 arcsec over an effective field of view of 37$\times$20 arcmin$\rm ^{2}$.
It is optimized to observe in a narrow wavelength range, $\rm 199-213~nm$.
This wavelength range corresponds to the `sweet spot' of dioxygen and ozone atmospheric absorption. 
FIREBall was launched in September 2018 from Fort Sumner, New Mexico, targetting Ly$\rm \alpha$ emission from z$\sim$0.7 galaxies, OVI emission from z$\sim$1 galaxies and CIV emission from z$\sim$0.3 galaxies.
We summarize the relevant characteristics of the instrument in Table \ref{tab:fireballspecifications}.
FIREBall is pathfinder experiment for a more ambitious project, ISTOS (PI: C. Martin, \citealt{Martin2014ISTOS}), a UV IFS satellite to be proposed to NASA.\\

\begin{table*}
\caption{FIREBall-2 Instrument specifications: We summarize here the critical characteristics of the FIREBall-2 instrument. These are also the instrument model parameters that directly impact our SNR analysis.}
\label{tab:fireballspecifications}
\begin{tabular}{ll}
\hline
Parameter  & Value \\
\hline
spectral resolution &$\sim$ 2000 $\lambda / \delta\lambda$ \\
FWHM & $\sim$ 5-6 arcsec \\
effective field of view & 37 x 20 arcmin$^2$ \\ 
wavelength range & 199-213 nm \\
diameter of mirror & 1m\\
number of objects observable per night & $\sim$ 200-300 with 2h exposure \\
sky background & 500 $\rm photons/s/cm^{2}/sr/$\AA \\
acquisition time per field & 2 hours\\
dark current & 0.036 $\rm e^{-}/pixel/hour$\\
induced charge & 0.002 $\rm e^{-}/pixel/frame$\\
read noise & negligible in photon counting mode \\
detector effective QE & $\sim$55\% \\
total optical throughput & 13\% \\
atmospheric throughput & 55\%\\

\hline
\end{tabular}
\end{table*}

\subsection{Instrument Model}

In order to prepare for the upcoming data analysis of FIREBall-2, \citet{Mege2015} developed a code that simulates the end-to-end image reconstruction process along the optical path of the instrument. 
This code, coupled to ZEMAX, generates a set of Point Spread Functions (PSFs) from an optical model at any given field positions and wavelengths. 
These PSFs are then interpolated at any point (in the field and wavelength), giving access to fundamental optical properties (magnification matrix, optical throughput, optical distortion, spectral dispersion) derived from the optical mappings existing between the sky plane and the instrument's mask or detector plane. 
Secondly, it produces 2D images of the electronic map of a detector patch corresponding to the observation of a modeled emission line from a simulated galaxy. 
Since its implementation in 2015 the code has constantly been modified according to the changes and updated measurements on the FIREBall instrument itself.
We use the instrument specific values given in table \ref{tab:fireballspecifications} for our calculation with the instrument model.

In the following, we combine the emission prescription to the instrument model to perform an end-to-end analysis of the observation of the CGM of low-redshift galaxies with the multi object slit spectrograph of FIREBall-2.

\subsection{Predicted signal with FIREBall observations}

The total image is a Poisson realization of the additive contribution of the CGM emission, the galaxy disc line emission, the continuum of the galaxy ($\rm \widehat{GAL}$), the sky ($\rm \widehat{SKY}$), the dark current from the detector ($\rm \widehat{DARK}$) and an induced charge current from the detector ($\rm \widehat{CIC}$). 
We use estimators for these contributions.
When observing an emission line from a galaxy halo, there are two contributions: the galaxy itself and the CGM.
There is no physical motivated border between the two, so for our further analysis we will call the combination ``extended line emission'' ($\rm \widehat{ELE}$).
Consequently the ``MeasuredSignal'' is {the sum of all contributions mentioned above and the signal we are interested in is the following}:
\begin{equation}
\rm \widehat{ELE} = MeasuredSignal - \widehat{GAL} - \widehat{SKY} - \widehat{DARK} - \widehat{CIC}
\end{equation}
The dark current is known from the calibration of the detector to be $\widehat{\mathrm{DARK}}$ = 0.0036 $\rm e^{-}/px/hour$ at $-110^{\circ}$C and a negligible variance due to an estimate on a large number of pixels with respect to the dominant noise source which is the galaxy continuum. We first remove the dark from the signal.
{We then estimate the profile of the continuum from regions towards the end of the galaxy spectrum, which are free of emission lines, $\rm P_{x,1}$ and $\rm P_{x,2}$, by stacking the columns of pixels (without the dark) over $\sim$ 10 columns in the dispersion direction. 
$x$ and $\lambda$ give the spatial and spectral coordinate on the detector.
The resulting continuum estimate is then an interpolation via linear regression between the two regions:}
\begin{equation}
\rm \widehat{\mathrm{GAL}}_{x,\lambda} = (1- \alpha) P_{x,1} + \alpha P_{x,2}
\end{equation}
{$\rm \lambda_1$ and $\rm \lambda_2$ are the central wavelengths of each region and $\alpha = \dfrac{\lambda-\lambda_1}{\lambda_2 - \lambda} $.}
{The corresponding variance is:
\begin{equation}
\rm \sigma^2_{\widehat{\mathrm{GAL}}_{x,\lambda}} = (1 - \alpha)^2 \sigma^2_{P_{x,1}} + \alpha^2 \sigma^2_{P_{x,2}}
\end{equation}
}
Considering a Poissonian distribution for the photon noise, the variance on the measurement is computed as the image $\rm \sigma^2_{x,\lambda,meas} = d_{x,\lambda}$. We neglect the read out noise as we use the detector in counting mode.
The last contribution to our SNR estimation comes from the induced charge of the detector, which is assumed a noiseless constant, $\widehat{\mathrm{CIC}}$.
Any other sources of noise are considered negligible.
As the current pixel size on the detector oversamples the resolution, we need to consider the contribution of a detector area corresponding to the actual resolution element. We therefore compute a SNR per resolution element by convolving the continuum-subtracted signal (and the corresponding noise) with the estimator for the instrument PSF ($\widehat{\mathrm{PSF}}$), normalized by the maximum pixel value.
The SNR per resolution element then becomes:
\begin{equation}
\rm SNR^{PRE}_{x,\lambda} = \dfrac{((d - \widehat{\mathrm{DARK}} - \widehat{\mathrm{GAL}} - \widehat{\mathrm{CIC}}) \ast \widehat{\mathrm{PSF}})_{x,\lambda}}{\sqrt{((d+\sigma^2_{\widehat{\mathrm{GAL}}}) \ast \widehat{\mathrm{PSF}})_{x,\lambda}}}
\end{equation}
In order to mock real observations we need to add some noise.
Therefore we use a Poissonian realization of the analytic solution of the SNR.
From this we can then infer the maximum SNR per resolution element for any input object of the FIREBall-2 IMO.

\begin{figure}
\centering
\includegraphics[width=0.5\textwidth]{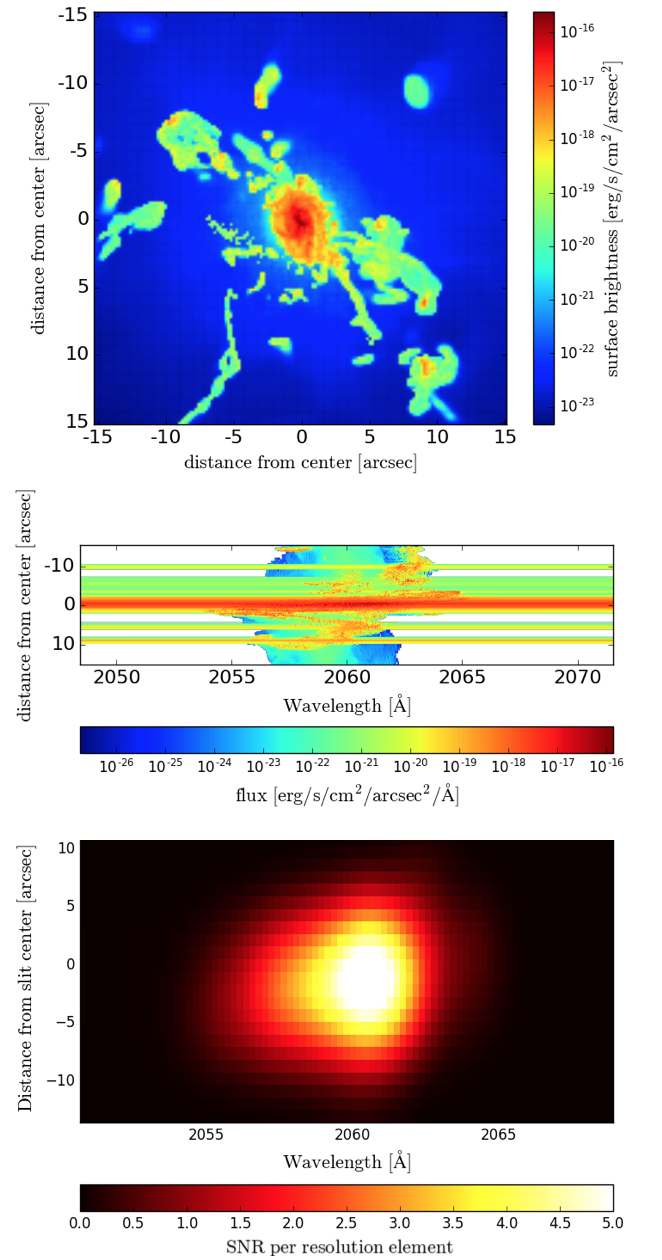}
\caption{Example of the FIREBall IMO input and output: Here we show the most massive Ly$\alpha$ halo input cube {at z=0.7}, illustrating the expected CGM flux from hydrodynamical RAMSES simulations, and output SNR. 
The upper panel shows the surface brightness of the simulated halo cube, the middle panel the projected spectrum of the same cube.
The lower panel shows the output SNR per resolution element for this simulated galaxy halo, after going through the IMO and data reduction, including removal of the galaxy continuum. The output is for a 2h observation.
} 
\label{im:snrexample}
\end{figure}

\subsection{Optimising FIREBall observing strategy}

\begin{figure}
\begin{center}
\includegraphics[width=.49\textwidth]{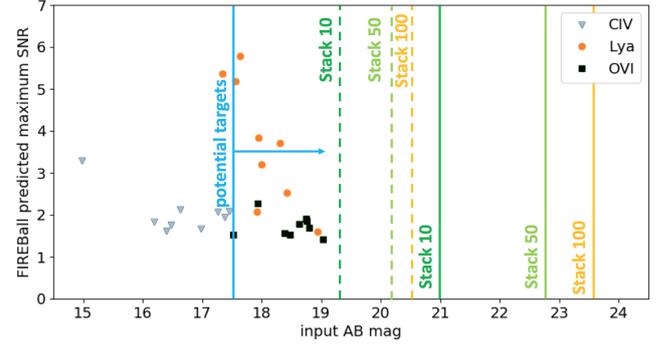}
\caption{Expected SNR for FIREBall targets: The different markers show the calculated SNR for the 10 most massive halos in the cosmological simulation at each redshift (0.3 for CIV, 0.7 for Ly$\alpha$ and 1.0 for OVI). 
The blue line shows the limit of the brightest potential targets for FIREBall-2. 
The typical target for FIREBall-2 will, however, be fainter than this upper limit. 
We considered two cases: The optimistic case (solid lines), where the ELE signal dominates the background and a pessimistic case (dashed lines), where the background dominates the observations.
The dark green lines shows the magnitude limit providing a SNR of 3 from stacking 10 Ly$\alpha$ sources in either case. The light green (yellow) lines shows the same limit from stacking 50 (100) sources.} 
\label{im:fireballresult}
\end{center}
\end{figure}

Now, we use our simulated halos as input into the IMO and perform the above described SNR calculation {for a 2h exposure}. 
From the results we then estimate the SNR of potential FIREBall targets.

For the SNR calculation, we choose the 10 most massive halos at the redshifts for CIV, Ly$\alpha$ and OVI (0.3, 0.7 and 1.0, respectively).
The properties of these halos are given in Table \ref{tab:halopropsfireball} {in the appendix}.
We use the most massive halos because they have the highest resolution in the AMR simulation.
The most massive Ly$\alpha$ halo is shown in the upper two panels of Figure \ref{im:snrexample}. 

We input those chosen 30 simulated galaxy halos into the FIREBall IMO and determine their SNR map with the prescription given above.
In the lower panel of Figure \ref{im:snrexample}, we show an example of the SNR map after data reduction for the most massive Ly$\alpha$ halo.
From the SNR map, we determine the maximum SNR and plot it against the NUV AB magnitude of the stellar continuum of the input halo (scatter points in Figure  \ref{im:fireballresult}).

By relating the maximum SNR of the ELE to the NUV magnitude, we can compare the simulated halos to actual galaxies and estimate which NUV magnitude corresponds to which maximum SNR given the emission line and redshift.
Since we chose the most massive halos from the simulation - which are supposedly also {some of} the brightest - we extrapolate from our results to lower magnitudes.
From the extrapolation we determine how many galaxies have to be stacked to give a reasonable SNR.

We present our simple estimate in the following:
\begin{equation}
\rm
SNR= \dfrac{S}{N} = \dfrac{S}{\sqrt{S+B}}
\end{equation}
Here we assumed all background components to be inside B.
Given the many uncertainties in determining the necessary parameters to calculate B, we simplify our analysis by considering the two extreme cases, each assumed to be valid in our complete magnitude range: One where S $>>$ B, which is optimistic (case1), and another one where S $<<$ B, which is a pessimistic case (case2).
This gives us the following approximations for SNR:
\begin{equation}
\rm
SNR (case1) = \dfrac{S}{\sqrt{S+B}} \approx \sqrt{S}
\end{equation}
\begin{equation}
\rm
SNR (case2) = \dfrac{S}{\sqrt{S+B}} \approx \dfrac{S}{\sqrt{B}} \propto S
\end{equation}
We also assume the ELE flux to be proportional to the total galaxy flux and relate the magnitude/flux of the galaxy to signal from the ELE:
$\rm F \propto S$.
\begin{equation}
\rm
m_1-m_2 = -2.5 log_{10}\left(\dfrac{F_1}{F_2}\right) = -2.5 log_{10}\left(\dfrac{S_1}{S_2}\right) 
\end{equation}
{Now we can relate the SNR for both cases to the difference in magnitudes.
We assume B to be constant in case2.
Given our results that the ELE from a galaxy with a continuum magnitude of 17.5 results in a SNR for Ly$\alpha$, we extrapolate to lower magnitudes with the following expression:}
\begin{equation}
\label{eq:snr1}
\rm
SNR (case1) = 10^{\left(\dfrac{17.5-m}{5}\right)} \times 5
\end{equation}
\begin{equation}
\label{eq:snr2}
\rm
SNR (case2) = 10^{\left(\dfrac{17.5-m}{2.5}\right)} \times 5
\end{equation}
{A table with specific values for given input magnitudes is given in the appendix (Table  \ref{tab:snrresultslya})}.
From these results we estimate that we will get good results for single galaxies at magnitudes up to NUV$\approx$18, even in the pessimistic case.
For fainter galaxies (NUV$>$19) we will have to stack single observations to reach a good SNR (3) in either case.
We assume to stack targets that give the same mean SNR individually and can thereby estimate the SNR that such a stack would give, at a given magnitude/SNR:

\begin{equation}
\rm
SNR_{stack} = \sqrt{Number\ of\ Objects} \times SNR_{individual\ object}
\end{equation}

Column 3 of Table \ref{tab:snrresultslya} gives the number of targets which need to be stacked at a given magnitude to reach the desired SNR of 3 for both cases.
In Figure \ref{im:fireballresult} we show {our results as} lines for the magnitude limits providing a SNR of 3 by stacking 10, 50 or 100 galaxies, where the dashed lines represent the pessimistic case and the solid lines the optimistic case.
The ELE SNR for OVI and CIV is found to be low even for sources that are bright in UV continuum.

\subsection{Target selection}
Based on these findings, we optimized the target selection and observing strategy for the launch in September 2018.
We favor bright quasars and Ly$\alpha$ emitting galaxies over the metal lines and aim primarily at dense fields with groups of quasars and Ly$\alpha$ galaxies in order to boost the signal through feedback.

The typical galaxy that qualifies as a target for FIREBall has a mean NUV magnitude of  $\rm mag_{NUV}\sim$23-24.
Therefore we aim to observe as many targets as possible during the night, in order to perform a stacking analysis.
In order to maximize the number of targets, we prepared four fields which should ideally be observed in equal amounts of time, resulting in 2h per field for a 8h night observation.
{From our SNR results we know that for a 2h observation we can get a good SNR for the bright objects while the faint ones need to be stacked.}
Each target field consists of up to $\sim$80 targets that fall into the right redshift windows.
Wherever there was still space in the field (on the mask), we put also some metal line galaxies, to make the best use of the detector.

In addition to the four science fields that need to be prepared in advance for mask cutting, the instrument will be equipped with a single slit for more flexible observations of e.g. an additional bright quasar, since bright targets are the most promising for the FIREBall observations.

\subsection{Expectations from the FIREBall experiment}

We analyzed the possible detection of CGM faint emission - or extended line emission ELE - from low-redshift galaxies with the FIREBall-2 UV MOS. 
We used mock cubes of an emission model on the FIREBall instrument model reproducing the output of the FIREBall-2 detector. The two dimensional analysis of the signal indicates that the massive objects can be observed in Ly$\alpha$ at redshift z=0.67 within the time available for the balloon's flight.
This {shows} the need for future development for the satellite version of the instrument, ISTOS.
Our simulations indicate that with the current version of the instrument and flight-plan it will be challenging to detect the OVI and CIV emission lines (at redshift 1.0 and 0.3 respectively).

We also considered stacking in order to achieve observability and a good SNR for the ELE.
For this we reviewed the continuum magnitudes of galaxies that can be potential targets for FIREBall-2.
The brightest FIREBall-2 targets have magnitudes NUV$\sim$18.
Those, including quasars, will give an excellent SNR with a single observation.
For the fainter targets (19$<$NUV$<$21) we would need to stack 10-300 galaxies to reach a good SNR.
Galaxies fainter than NUV=21 - including the bulk of the FIREBall-2 targets with a mean $\rm NUV \sim$23-24 - will be challenging to observe even when stacking the signal, when assuming that the observations are dominated by the background.
In the other extreme case, where the observations are dominated by the signal of the object, we expect to obtain the desired SNR of 3 when stacking $\sim$100-300 galaxies down to NUV=24.
The real case will lie somewhere in-between those two extremes.

One remaining issue with the FIREBall-2 observations will be the separation of the CGM from the disc line emission. 
With the current spatial resolution we would need a highly luminous and extended CGM to resolve it separately from the disc. 
In case this is not possible, we will have to make assumptions on the ratio between Ly$\alpha$ disc emission and CGM emission and apply it to the total signal in order to estimate the CGM flux.

Future satellite missions like ISTOS \citep{Martin2014ISTOS} or LUVOIR \citep{France2017} will enable us to see the UV emission of galaxies at these redshifts at an even better SNR and thus will be able to spatially resolve the CGM.

\section{Optical and Near-Infrared observations with ELT/HARMONI}

The High Angular Resolution Monolithic Optical and Near-infrared Integral field spectrograph\footnote{http://www-astro.physics.ox.ac.uk/instr/HARMONI/} (HARMONI, PI: N. A. Thatte, \citealt{Thatte2014}) will be the integral field spectrograph (IFS) at the ESO Extremely Large Telescope\footnote{https://www.eso.org/sci/facilities/eelt/} (ELT).
HARMONI will be available with different flavors of Adaptive Optics (AO) systems.
It can be used without AO, with Laser Tomography AO (LTAO) or with Single Conjugate AO (SCAO).
The instrument will cover a wavelength range from 0.47 $\rm \mu m$ in the visible to 2.45 $\rm \mu m$ in the near-infrared.
There will be four different spatial scales available with associated fields-of-view.
For our purposes we will only consider the widest field-of-view, which has the biggest spaxels.
This coarser spatial resolution mode with spaxels of a size of 60 $\times$ 30 mas will have the largest field of view (6.42 $\times$ 9.12 arcsec), which is most appropriate to study the large extent of the CGM outside the host galaxies.
{HARMONI can achieve a spectral resolution} between R $\sim$ 3000 and R $\sim$ 20000, depending on the wavelength regime.
The ELT and HARMONI are planned to have first light in late 2024.

\subsection{The HARMONI instrument simulator (HSIM)}

\begin{figure}
\includegraphics[width=.5\textwidth]{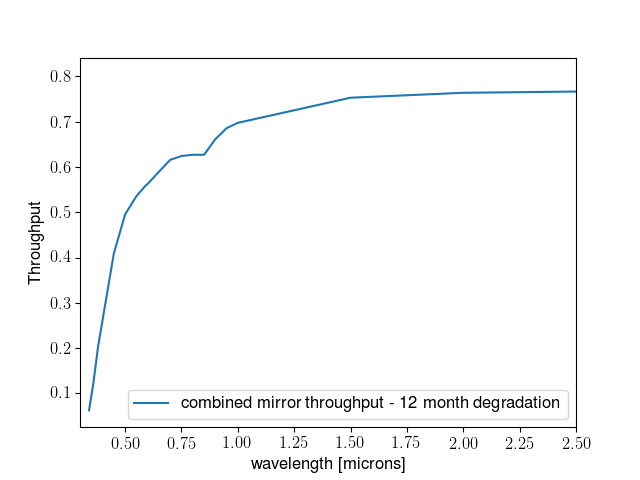}
\caption{Combined reflectivity of the six mirrors in the ELT for a 12 month degradation.
While the throughput is more than 70\% at NIR wavelengths the visible, and especially the blue wavelengths are heavily absorbed.
Curve taken from HSIM \citep{Zieleniewski2015} version 115. {It assumes a Gemini 4 Layer Ag coating for 5 of the six mirrors in the ELT and a fresh Al coating for one of them \citep{Boccas2004,Hass1965}.}
}
\label{im:reflectivity}
\end{figure}

In preparation of the science objectives with HARMONI, a simulation tool called HSIM\footnote{https://github.com/HARMONI-ELT/HSIM} has been developed \citep{Zieleniewski2015}.
It is an instrument model and calculates the observed signal and noise for a given input source, taking into account all instrumental and atmospheric effects. 
In particular, it takes into account the reflectivity of the telescope mirror coating and its degradation over time.
This is particularly essential for our science objective, as the coating shows poor performance in the blue.
At 5000 \AA\ an overall reflectivity of $\sim$ 50\% or less is expected (for reflections on the 6 mirrors of the telescope, see Figure \ref{im:reflectivity}), depending on the state of degradation.
HSIM returns a reduced mock observation of the input object.
We use version 115 of HSIM to determine the expected signal from the CGM using HARMONI.

We prepare three dimensional data cubes of simulated galaxy haloes in a similar way as for the FIREBall IMO.
For the observation simulations we use the V+R, Iz+J and H+K gratings, giving a spectral resolution of $\rm R=3100-3300$ and the coarse spaxel scale with pixels of 30 $\times$ 60 mas.
For adaptive optics we use the Laser Tomography Adaptive Optics (LTAO) which uses laser guide stars.
Given our setup (biggest spaxels), a tip-tilt star (TTS) free mode will provide a full sky coverage.
We set the Zenith seeing to 0.67 arcsec and the Zenith angle to 0 deg.
The telescope temperature is set to 280.5 K as the default temperature given by HSIM.

\subsection{Simulated input cubes} \label{inputcubes}

As for FIREBall, we use the post-processed galaxy halo simulations described {earlier}.
We consider Ly$\alpha$, OVI and CIV as potential tracers for CGM emission.
Additionally we use H$\alpha$ as a tracer for low redshift CGM.
Table \ref{tab:redshiftcoverage} gives an overview of the lines and their respective redshifts.
At each redshift we consider the most massive halo, because they have the highest resolution in the AMR RAMSES simulation.
These halos are investigated for the general CGM properties such as angular extent and luminosities (see section \ref{cgmevol}).

For the HSIM input and the estimation of flux-dependent SNR at different wavelengths we use the most massive halo at redshift 0.3.
The properties of this halo are given in the first line in Table \ref{tab:halopropsfireball}.
The line we choose in this halo is H$\alpha$.
We pick this halo because of its high resolution and gas-rich CGM. 
Defining an area of 0.6 $\times$ 0.6 arcsec$^2$ around a gas cloud in its CGM, we want to know how the SNR of the flux in this area changes with wavelength.
Therefore we shift the input cube's wavelength in steps of $\rm \Delta\lambda=0.05 \mu m$ to populate the spectral coverage of HARMONI.
For each of these cubes at different wavelengths, we also modify the flux by scaling by factors {ranging from 10$^{-4}$ to 10$^{4}$ in 9 log steps.}
Thereby we end up with 9 different input fluxes at each wavelength for which we measure the output SNRs.

\begin{figure}
\begin{center}
\includegraphics[scale=0.7]{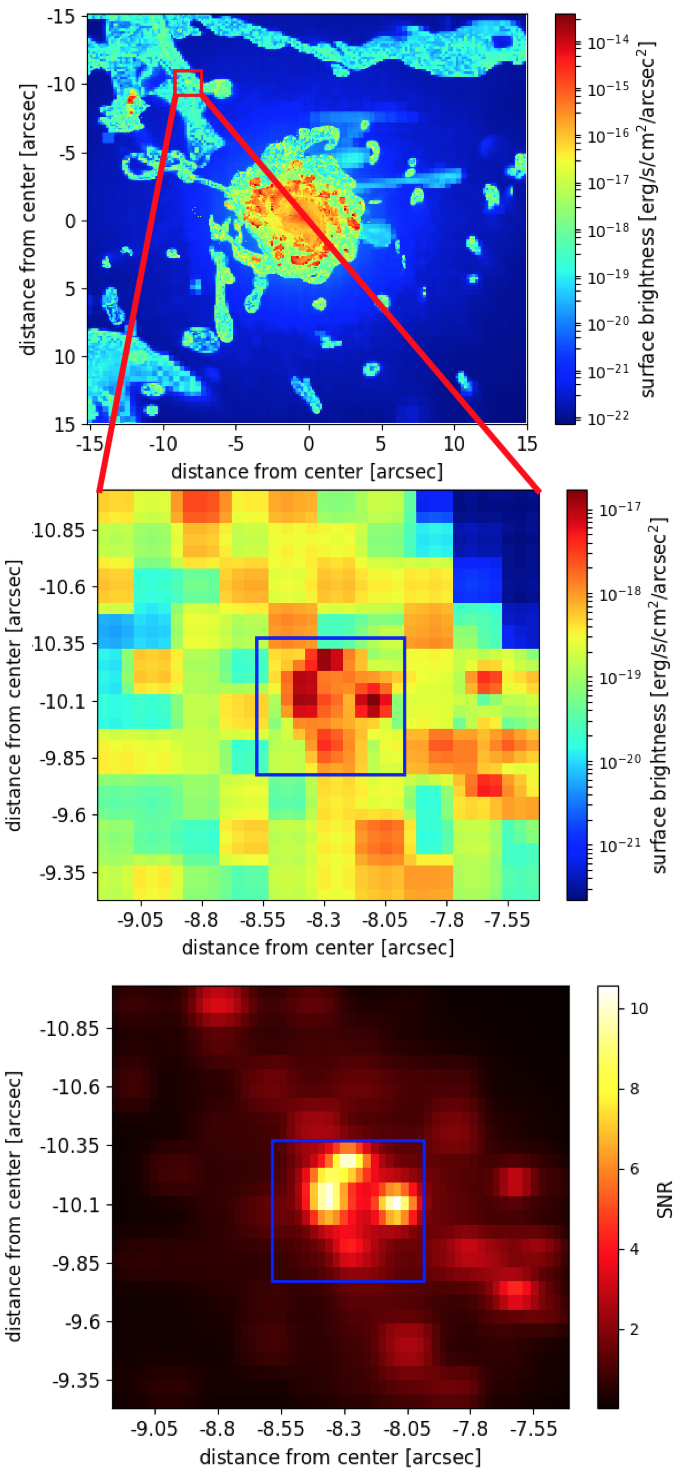}
\caption{Example of HSIM input and output: 
The upper panel shows the H$\alpha$ emission in the most massive galaxy halo at redshift 0.3 in our RAMSES AMR hydrodynamical simulations. 
From this cube - in order to save computation time - we extract a smaller data cube from one of the filaments in the CGM region around the central galaxy (middle panel).
The cube was shifted to different wavelengths in order to populate the spectral range of HARMONI.
In the bottom panel we show the output SNR for the input cube at 1.32 microns in the Iz+J grating for an integration time of 5 hours.
The blue box in the lower two panels gives the region of 0.6 $\times$ 0.6 arcsec, which is considered for the analysis.
}
\label{im:hsimexample}
\end{center}
\end{figure}

In Figure \ref{im:hsimexample}, we show one example of input and output of HSIM.
The upper two panels show the {H$\alpha$} surface brightness map of the most massive halo at redshift 0.3. 
This halo was modified according to the above description and shifted in wavelength, so that the lower panel shows the SNR for the cube at a wavelength of 1.32 microns.

\subsection{CGM evolution and observability} \label{cgmevol}

\begin{table}
\centering
\caption{Redshift coverage of different lines with HARMONI (0.47 - 2.45 $\rm \mu m$)}
\label{tab:redshiftcoverage}   
\begin{tabular}{|c|c|c|}
\hline
Line & redshift range & chosen redshift for simulations \\ \hline
H$\alpha$ & 0.0 - 2.7 & 0.3, 1, 2 \\
Ly$\alpha$ & 2.9 - 19 & 4, 6, 10 \\
OVI & 3.5 - 22 & 4 \\
CIV & 2.0 - 14 & 3 \\
\hline
\end{tabular}
\end{table}

The extended wavelength range of HARMONI will allow us to observe different CGM tracers at various redshifts (see Table \ref{tab:redshiftcoverage}).
We know that at low redshifts, the CGM can reach out to several hundreds of kpc \citep{Tumlinson2017}.
The maximum field-of-view of HARMONI is 9$\times$6 arcsec$^2$.
At $z=0.3$, one arcsec corresponds to $\sim$4.5 kpc.
It will not be possible to map the full CGM region in one exposure with HARMONI at this redshift.
Due to cosmic evolution, angular scale will be smallest between $z=1$ and $z=2$.
Beyond $z\sim$ 1-2, the CGM will appear even smaller due to the early stages of galaxy evolution itself but also fainter due to redshift effects.
Therefore observations of galaxy halos at $z\sim$ 1-2 will be optimal to map the CGM.
At these redshifts the largest field-of-view of HARMONI of 9 $\times$ 6 arcsec will correspond to $\sim$ 75 $\times$ 50 kpc.
The virial radii of galaxy halos at those redshifts can stretch out to $\sim$ 200-300 kpc, so the majority of the surroundings of a galaxy could be covered with $\sim$ 4 neighboring exposures.
For Ly$\alpha$ which is the brightest line at any redshift, we conclude that the optimal redshift for CGM observations is z $\gtrsim$ 3, because from redshift 3 the virial radii of galaxies will typically be $<$ 70 kpc and it will be possible to capture the CGM in a single exposure.

\begin{figure*}
\begin{center}
\includegraphics[scale=0.5]{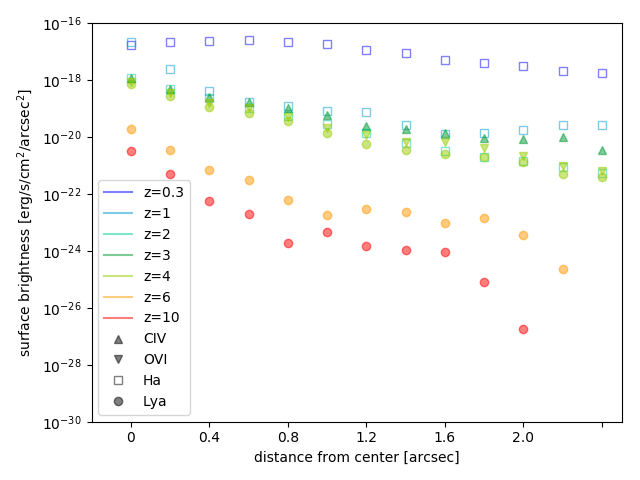}
\includegraphics[scale=0.5]{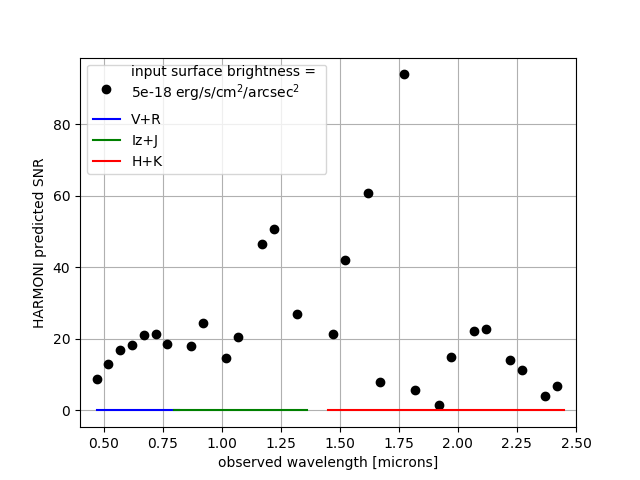}
\caption{Left panel: Radial profiles of simulated galaxy halos: We show the radial surface brightness profiles of galaxies from our cosmological RAMSES AMR simulations for different ions at various redshifts.
On the x-axis we plot the distance from the center of the galaxy halo in units of arcsec, which are independent of redshift and therefore translate to different physical sizes for a given redshift.
The points give the average surface brightness (on y-axis) of the halo at a given distance.
As expected, the overall surface brightness decreases with redshift.
We also notice that Ly$\alpha$ emission at redshift 4 is comparable to H$\alpha$ emission at z=2 and metal lines at redshifts 3 and 4.
Right panel: Wavelength dependence of output SNR in HARMONI: All SNR points are calculated for a given input surface brightness of $\rm 5e-18 erg/s/cm^{2}/arcsec^{2}$. The ranges of the V+R, Iz+J and H+K gratings are shown in blue, green and red, respectively.
We see a steady increase of SNR in the blue which traces the throughput of the telescope, given a coating on the mirror that absorbs heavily at these wavelengths.
The throughput in the NIR is higher and more steady than in the visible, but subject to atmospheric absorption lines and OH emission lines (especially in the H+K band) so that the SNR varies strongly with the precise wavelength.
}
\label{im:radialprofile+wldep}
\end{center}
\end{figure*}

To illustrate the flux evolution within a given angular size of 2 arcsec - corresponding to $\sim$ 9-17 kpc, depending on the redshift - we plotted in Figure \ref{im:radialprofile+wldep} the radial profiles of the most massive halos at the redshifts and for the lines given in Table \ref{tab:redshiftcoverage}.
Naturally, the low redshift halo at z=0.3 gives the brightest flux profile.
We also see the steep drop in luminosity for Ly$\alpha$ between z=4, 6 and 10. 
Ly$\alpha$, being the brightest emission line in any galaxy halo, also shows comparable fluxes at z=4 to H$\alpha$ emission at z=2 and metal lines at redshifts 3 and 4.

\subsection{Predicted signal from HARMONI observations}

We run HSIM for a grid of wavelengths and fluxes as described in section \ref{inputcubes}.
The observing time is chosen to be 5 hours with 5 integrations of 3600 seconds each. 
To optimize the observation setup, we tested different exposure settings in the V+R grating and found that longer integration times and fewer exposures give a better SNR than choosing more exposures with shorter integration times.
Specifically for the fixed 5 hours, we find a 7\% increase of the SNR when choosing 5 $\times$ 3600 seconds over 20 $\times$ 900 seconds.

After running HSIM for each of our input cubes, we determine the corresponding SNR of the output.
In the input cubes we have chosen an area of 0.6$\times$0.6 arcsec$^2$ around a gas clump.
In the output cubes, we derive the SNR of the same area by binning 20$\times$10 pixels in the HSIM output.

First, we consider the input cubes with the original input flux from the cosmological simulations.
We investigate the wavelength dependence of the output SNR for a given flux (Figure \ref{im:radialprofile+wldep}).
While the SNR increases with wavelength in the visible, just as expected from the telescope's throughput, there is a large scatter of SNRs at larger wavelengths.
At these wavelengths the instrument's throughput is better than in the visible, resulting in high SNRs.
But there are also numerous atmospheric absorption lines and OH emission lines which corrupt the observation and lead to low SNRs.
By choosing random input wavelengths, some regions are affected by the atmosphere and some others are not (e.g. 1.77$ \rm \mu m$) and give an indication of the range of SNR in HARMONI NIR observations.

\begin{figure}
\begin{center}
\includegraphics[scale=0.5]{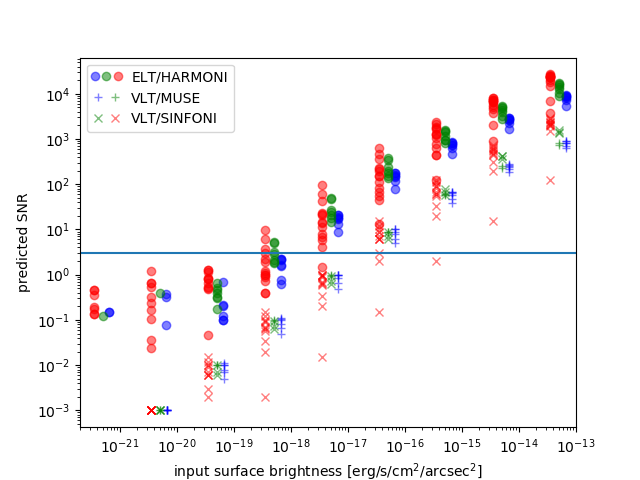}
\caption{Expected signal-to-noise for different input fluxes and comparison to existing IFSs (MUSE and SINFONI) for a 5 hour observation.
We plot for each pair of surface brightness and wavelength the output SNR. The circles give the SNR for HARMONI, the '+' the SNR for MUSE and the '$\times$' the SNR for SINFONI.
We group wavelengths that would fall into the V+R, Iz+J and H+K band of HARMONI with blue, green and red colors, respectively.
For display purposes, we offset the points in the different wavelength bands in x-direction.
The green points are at the original surface brightness, the red and blue ones have been shifted to the left and right, respectively.
We also show the SNR=3 limit, which is the minimum SNR that should be achieved for kinematic modelling \citep{Bouche2015,Peroux2017}.
We find a significant increase of SNR in HARMONI compared to both MUSE and SINFONI. 
The ratios between HARMONI and MUSE SNR are shown in Figure \ref{im:musecomp+sinfonicomp}, the ratios between HARMONI and SINFONI SNR are shown in Figure \ref{im:musecomp+sinfonicomp}.
}
\label{im:snr}
\end{center}
\end{figure}

We also use the flux-modified input cubes at each wavelength and determine how the SNR changes with both input flux and wavelength.
In Figure \ref{im:snr} we show the result of this computation.
We plot the output SNR against input flux.
The color-coding corresponds to the chosen grating.
For each flux there is a spread in SNRs for each grating because of the different wavelengths we analysed within each grating.

\subsection{Comparison with current IFSs at optical/NIR wavelengths}

To quantify the gain from ELT/HARMONI over current state-of-the-art IFSs such as MUSE and SINFONI on the VLT, we compare our findings for the expected SNR of the CGM to the SNR we expect to have with these instruments for the same given flux of the CGM.
To do so we use the online Exposure Time Calculators (ETCs) for both MUSE\footnote{http://www.eso.org/observing/etc/bin/gen/form?INS.MODE= swspectr+INS.NAME=MUSE} and SINFONI\footnote{https://www.eso.org/observing/etc/bin/gen/form?INS.NAME= SINFONI+INS.MODE=swspectr}.

\subsubsection{Optical IFS VLT/MUSE}

\begin{figure*}
\begin{center}
\includegraphics[scale=0.5]{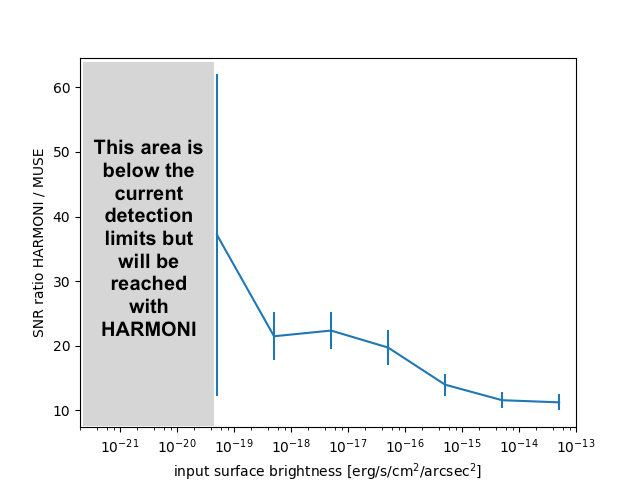}
\includegraphics[scale=0.5]{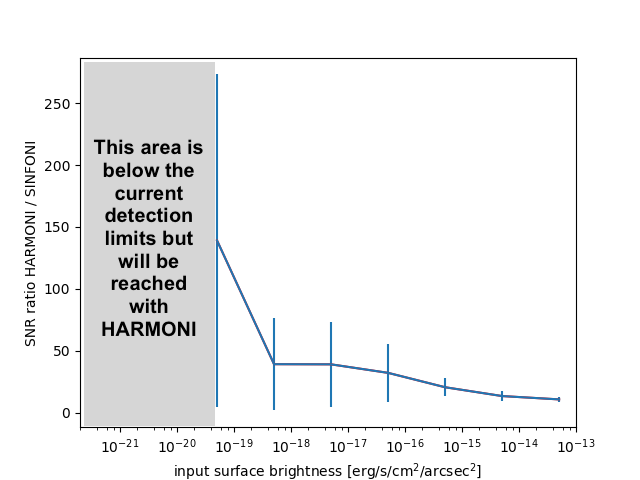}
\caption{Left panel: SNR ratio between HARMONI and MUSE: For each wavelength and input flux, we determine the ratio between SNR in HARMONI and SNR in MUSE.
For each value in input flux we compute the mean and standard deviation of these ratios at all wavelengths.
We find an increase of SNR in HARMONI over MUSE at all wavelengths and fluxes of at least an order of magnitude.
We predict a steepening increase of this ratio towards low surface brightness. 
The standard deviation increases significantly at low surface brightness as the SNR in the MUSE data becomes very low and a large spread of SNR ratios becomes possible.
This means that faint diffuse emission from the CGM which is currently not detectable in even the deepest MUSE observations will become observable in HARMONI.
{Right panel: same as left panel but for SINFONI}.
}
\label{im:musecomp+sinfonicomp}
\end{center}
\end{figure*}

MUSE is an IFS at the VLT and covers a wavelength range from 480 nm to 930 nm. 
It has been successfully used in tracing extended Ly$\alpha$ emission around galaxies (e.g. \citealt{Wisotzki2016}).
We use version P102.7 of the MUSE ETC to calculate the expected SNR of the same line emission as for HARMONI with HSIM.
For the source emission we assume an extended source with 1.13 arcsec diameter (to result in 1 arcsec$^2$ area) and single line emission at the same wavelengths and fluxes as for HARMONI.
We use the Wide Field Mode without AO and a spatial binning of 3$\times$3 pixels to reach the same area of 0.6$\times$0.6 arcsec$^2$ as in the simulation.
We assume an airmass of 1.5, moon FLI of 0.5 and seeing of 0.67 arcsec.
The exposure time is also set to the same amount as for HSIM: 5 $\times$ 3600 seconds.
Our results for the obtained SNR in MUSE is plotted in Fig. \ref{im:snr} with crosses (+) and in colors blue and green, corresponding to the wavelength ranges of the HARMONI gratings.

We find an overall increase of a factor $\sim$20 in SNR for HARMONI observations over MUSE observations.
The ratios of SNRs (HARMONI/MUSE) are shown in Figure \ref{im:musecomp+sinfonicomp}.
While there is generally an increase of more than one order of magnitude at fluxes $\rm > 10^{-18} erg/s/cm^{2}/arcsec^{2}$, we also find that small fluxes that would have been undetected even in deep observations with MUSE (detection limit for emission lines in $\sim$ 20-30h MUSE observations is $\rm 2.8 - 5.5 \times 10^{-20} erg/s/cm^{2}/arcsec^{2}$/\AA\ in the most ideal cases \citep{Leclercq2017} generally it is around $\rm \sim 1 \times 10^{-19} erg/s/cm^{2}/arcsec^{2}$, \citealt{Wisotzki2016,Bacon2017,Wisotzki2018}) will become observable in HARMONI.
Thus, HARMONI will enable new CGM science.
This will mark the next step in CGM studies, where we are limited by the current instrument sensitivities.
We will be able to map the CGM and get measurements on its extent and clumpiness.
Notwithstanding the ELT's mirror coatings, which is suboptimal at visible wavelengths, the increase in collecting area means that photon-starved science cases would benefit from the ELT even at visible wavelengths.

\subsubsection{NIR IFS VLT/SINFONI}

SINFONI, is the NIR IFS at the VLT and operating in the near infrared from 1.1 microns to 2.45 microns. 
We vary the input flux and wavelength and determine the output SNR with the SINFONI ETC version P102.7.
We assume an extended source, with an area of 0.36 $\rm arcsec^2$, because the output SNR in the SINFONI ETC is given for the entire source size.
The AO and sky conditions are the same as for the MUSE ETC.
The angular resolution scale is set to 250 milliarcsec to get the maximum sensitivity and biggest FOV and we use the J, H and K-band grating for the respective wavelengths.
We assume a total exposure time of 5 hours but due to the sky variations at NIR wavelengths in SINFONI we split it into 20$\times$900 seconds.
Our results are again plotted in Fig. \ref{im:snr} with an '$\times$' and in red and green, corresponding to the respective gratings in HARMONI.

We find an increase of at least a factor $\sim$15 for HARMONI observations over SINFONI observations, with a mean between a factor 15 to 100.
We plotted the ratios for all fluxes in Figure \ref{im:musecomp+sinfonicomp}: 
The expected SNR increases at all redshifts and the small fluxes which were previously not observable will become detectable with HARMONI.
{SINFONI will be decommissioned in 2019\footnote{https://www.eso.org/sci/facilities/paranal/cfp/cfp102/foreseen-changes.html} and replaced in 2020 by ERIS\footnote{https://www.eso.org/sci/facilities/develop/instruments/eris.html}.
By the mid 2020s, when the ELT will be available, HARMONI will make a more than suitable replacement for the only NIR IFS at large ESO telescopes.}

\subsection{Future CGM studies with HARMONI}

As we have shown in Figure \ref{im:snr}, we expect HARMONI to detect at least one order of magnitude smaller fluxes than previously possible and we will be able to detect diffuse emission which is an order of magnitude fainter in surface brightness than the faintest detectable emissions discovered by MUSE and SINFONI.
This means that ELT/HARMONI will be {well} suited for photon starved science cases such as the faint diffuse emission from the CGM.
Even though the mirror coating of the ELT has suboptimal reflectivity at visible wavelengths, the telescope's large collecting area provides an improved signal with respect to VLT/MUSE observations.

\section{Conclusion}

We have dealt with the complex question of CGM faint emission modelling in order to produce realistic data cubes that can be used for observability predictions of the CGM with upcoming instruments.
We have used a state-of-the-art high resolution hydrodynamical cosmological RAMSES simulation to extract different massive halos ($\rm 10^{13} M_{\odot}$). Using a photo-ionization code, we modeled different line emissivities considering the UVB fluorescence and the gravitational cooling of the gas. We also considered the stellar contribution to the gas fluorescence in the case of Ly$\alpha$ photons and we derived the level of the UV continuum in those wavelengths after attenuation by the ISM dust.
{Our simulations include feedback from supernova explosions, modeled such that it creates artificially hot 'delayed-cooling' cells. In our model we exclude those cells in order to stay conservative in terms of total luminosity.}

We find our simulations to be in good agreement with low-redshift observations from GALEX \citep{Deharveng2008,Wold2014,Wold2017} for Ly$\rm \alpha$ escape fractions between 0.1\% and 1\%. 
Moving to higher redshifts (z=4.0 and z=2.33), our CGM Ly$\alpha$ emission model agrees well with the observational data provided we use a lower Ly$\alpha$ escape fraction than is usually inferred from observations. This effect might originate from the stacking of a large number of objects in the z=2.33 case.
Using our simulations, we can create simulated data cubes of mock observations with two spatial axes and one spectral axis.

We have {also} investigated the expected signals from CGM emission with two upcoming instruments: FIREBall-2 and HARMONI on the ELT.
We used {the} simulated halos as input into the respective instrument models of FIREBall-2 and HARMONI.
From these simulations we get an estimate of the signal that faint diffuse emission gives in observations with each of these instruments.
Those results give the base for target selection and observing strategies.

Our simulations and analysis have given us a basis on which targets to select - focusing on Ly$\alpha$ rather than the metal lines CIV and OVI.
While observations of individual objects will be challenging and probably only bright UV objects like quasars provide a high SNR, the instrument is designed such that it will be able to observe several hundreds of galaxies in one night. 
Stacking the signal of several hundred galaxies will be the way of analysing the FIREBall-2 data to gain new insights into extended Ly$\alpha$ emission at low redshifts.
FIREBall-2 was launched in September 2018 and observed the low-z CGM for the first time.
The data analysis of FIREBall-2 data is currently ongoing.

HARMONI, which has successfully passed the Preliminary Design Review (PDR), is planned for first light in late 2024.
The instrument design allows for a reliable instrument model which we use to prepare future CGM observations.
HARMONI will be a visible and NIR IFS and able to target different CGM tracers at various redshifts.
We have investigated the SNR expected for various input fluxes at different wavelengths and compared to the existing IFSs MUSE and SINFONI on the VLT.
We find an increase of $\sim$ 20 times better SNR with HARMONI compared to the current instruments.
This will allow us to reach one order of magnitude fainter surface brightness of faint diffuse emission than current facilities and will enable CGM studies.
Going to higher redshifts ($z \sim 1-2$) will allow us to map larger areas and in combination with the less evolved galaxies and surroundings, it will be possible to map galaxies with their entire CGM.
Therefore we conclude on a `sweet spot' at redshift $\sim$1-2 for general CGM observations and Ly$\alpha$ to be well observable at z = 3 $-$ 4.
HARMONI will enter a regime of low surface brightness which is not attainable with current facilities.
Also, while MUSE has a bigger field-of-view than HARMONI and is able to detect more galaxies in one exposure, HARMONI will allow us to reach lower surface brightness (SB> $\sim$ $\rm 10^{-19} - 10^{-20} erg/s/cm^{2}/arcsec^{2}$) in a 5 hour exposure.

Overall, the future looks promising for CGM studies with many upcoming new instruments, such as ISTOS \citep{Martin2014ISTOS} or LUVOIR Ultraviolet Multi-Object Spectrograph (LUMOS, \citealt{France2017}). 
Apart from the instruments that we have studied in this work it will be important to assess how space-based X-ray missions like ATHENA will shed new light onto the hot gas content of galaxy halos and address the missing baryon problem in the low-redshift universe \citep{Nicastro2018}.

\section*{Acknowledgements}

RA thanks CNRS and CNES (Centre National d'Etudes Spatiales) as well as ESO for support for her PhD.
CP is grateful to the ESO science visitor programme and the DFG cluster of excellence 'Origin and Structure of the Universe' for support. 
CP thanks the Alexander von Humboldt Foundation for the granting of a Bessel Research Award held at MPA. 
NT, LR, MPS \& SZ acknowledge support from the Science and Technology Facilities Council (grants ST/N002717/1 and ST/M007650/1), as part of the UK E-ELT Programme at the University of Oxford.
The simulations in this work are part of the BINGO! project.
This work was granted access to the HPC resources of CINES under the allocation 2015-042287 made by GENCI.
Images created with glnemo2 (https://projets.lam.fr/projects/glnemo2), a 3D visualisation program for nbody data, developed by Jean-Charles Lambert at CeSAM-LAM.
We thank J.-C. Lambert for GLNemo2 related discussions.
We thank ESO for the publicly available Exposure Time Calculators for MUSE and SINFONI.
RA thanks J{\'e}r{\'e}my Fensch and Nicolas Guillard for useful discussions.

\bibliographystyle{mnras}
\bibliography{library}


\appendix

\section{Specifics of the high resolution simulation}

\begin{figure}
\centering 
\includegraphics[width=.35\textwidth]{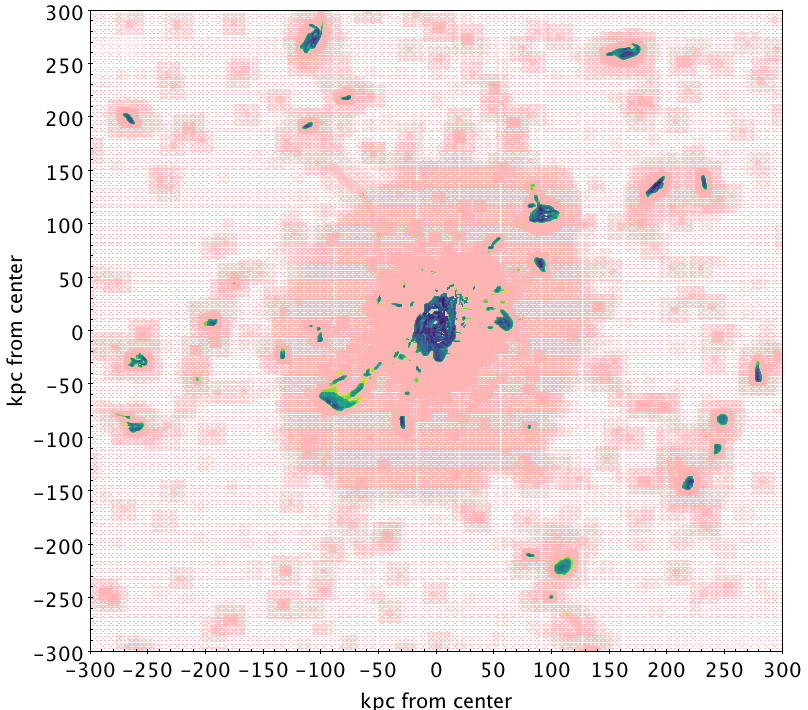}
\caption[Discs in the Simulation]{Example for the emergence of discs in our simulation with the green-blue color indicating the density and temperature combinations that were unreachable with the previous, low-resolution simulation. The redshift of this snapshot is 0.7. {The halo shown here is the one we discuss in section 3 and also for the FIREBall-2 analysis.}}
\label{fig:discs}
\end{figure}

\begin{figure}
\centering
\includegraphics[width=.35\textwidth]{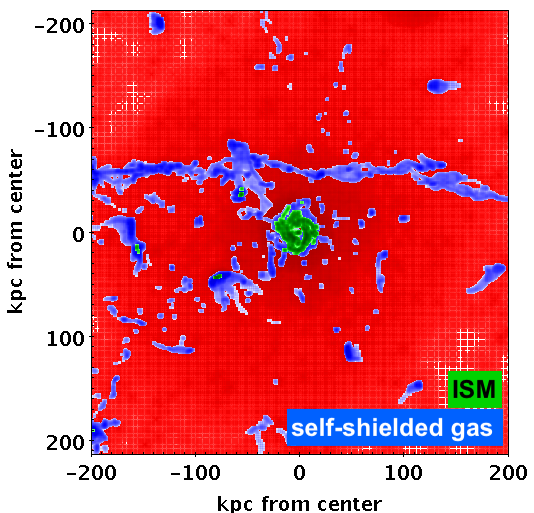}
\caption[Self-shielded gas and ISM]{Example of the distribution of gas cells that fall into our definition of self-shielded (blue) and ISM (green). For the chosen cuts on each phase, see Figure \ref{emissivityfromhalo}.{The halo shown here is the one we use for the HARMONI analysis at z=0.3.}}
\label{fig:ismplot}
\end{figure}

\begin{figure}
\centering
\includegraphics[width=.35\textwidth]{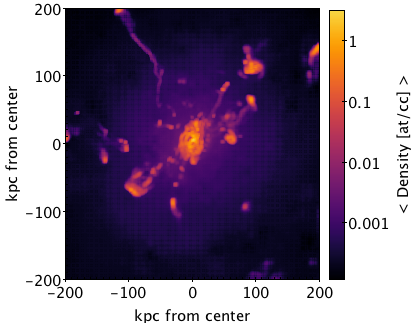}
\includegraphics[width=.35\textwidth]{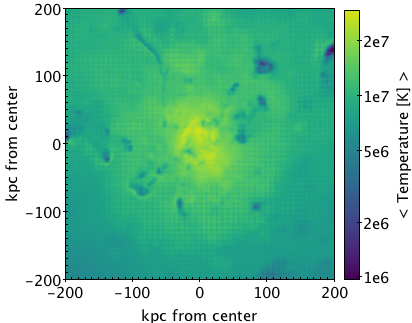}
\caption[NH and T map of galaxy halo at z=0.7]{The upper panel shows the total hydrogen density map for the most massive halo at z=0.7. The lower panel shows the temperature map of the same halo. {The halo shown here is the one we discuss in section 3 and also for the FIREBall-2 analysis.}}
\label{fig:nhimap-tmap}
\end{figure}

\begin{figure}
\begin{center}
\includegraphics[width=.35\textwidth]{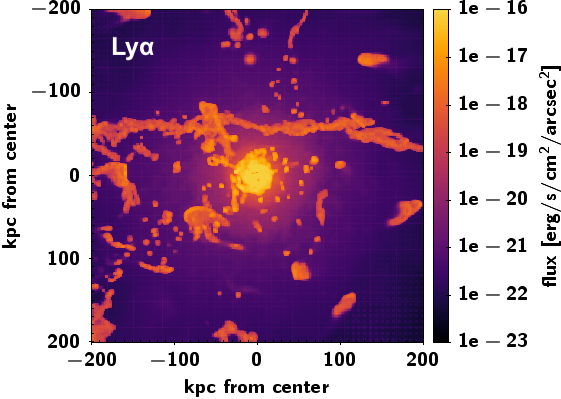}
\includegraphics[width=.35\textwidth]{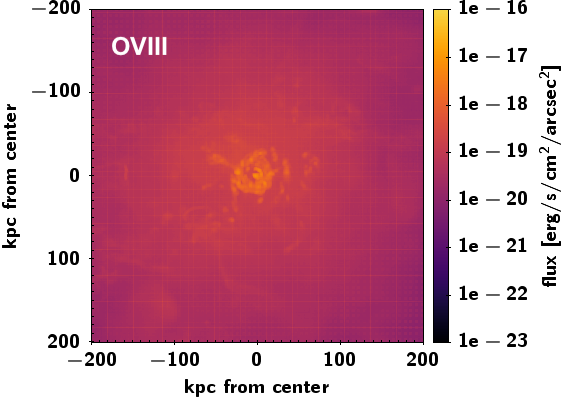}
\caption[Ly$\alpha$ and OVIII emission maps]{Ly$\alpha$ and OVIII emission maps at z=0.3. While the `cool' hydrogen gas emits strongly from clumps in the halo, the X-ray emission of e.g. OVIII is more homogeneous throughout the halo. {The halo shown here is the one we use for the HARMONI analysis at z=0.3.}} 
\label{im:xrayvsuv}
\end{center}
\end{figure}

{
Figure \ref{fig:discs} shows in green-blue the spatial distribution of new, high-resolution gas cells, which were not accessible in \cite{Frank2012}. They are mainly associated with ISM-like gas.
Figure \ref{fig:ismplot} shows the spatial distribution of the ISM and the self-shielded gas according to our definition, given in the main text.
Figure \ref{fig:nhimap-tmap} gives maps of the typical densities and temperatures in our {most massive} halos.
Figure \ref{im:xrayvsuv} shows the Ly$\alpha$ and OVIII emission maps in the {most massive} halo at z=0.3.
}

\section{Additional information on the FIREBall-2 SNR}

{
Table \ref{tab:halopropsfireball} gives the properties of all the 30 haloes that were used as input to the FIREBall-2 instrument model.
Table \ref{tab:snrresultslya} gives the SNR for a halo with a given input magnitude and the number of objects to be stacked at this magnitude to reach and SNR of 3.
}

\begin{table}
\centering
\begin{tabular}{ccccc}
\hline
line	&	redshift	&	$\rm M_{DM}$	&	$\rm M_{\star}$	&	SFR		\\
	&		&	[$\rm M_{\odot}$]	&	[$\rm M_{\odot}$]	&	[$\rm M_{\odot}/yr$]		\\
		\hline
CIV	&	0.3	&	2.83e+13	&	3.37e+12	&	333		\\
CIV	&	0.3	&	6.64e+12	&	9.98e+11	&	60		\\
CIV	&	0.3	&	3.51e+12	&	5.10e+11	&	37		\\
CIV	&	0.3	&	1.64e+12	&	2.34e+11	&	26		\\
CIV	&	0.3	&	1.17e+12	&	1.78e+11	&	14		\\
CIV	&	0.3	&	1.03e+12	&	1.40e+11	&	10		\\
CIV	&	0.3	&	6.18e+11	&	7.21e+10	&	7		\\
CIV	&	0.3	&	6.04e+11	&	1.82e+10	&	11		\\
CIV	&	0.3	&	4.53e+11	&	4.95e+10	&	7		\\
CIV	&	0.3	&	4.20e+11	&	4.09e+10	&	4		\\
										
Ly$\alpha$	&	0.7	&	1.36e+13	&	1.11e+12	&	246		\\
Ly$\alpha$	&	0.7	&	5.96e+12	&	7.91e+11	&	169		\\
Ly$\alpha$	&	0.7	&	3.07e+12	&	4.02e+11	&	82		\\
Ly$\alpha$	&	0.7	&	2.64e+12	&	3.01e+11	&	96		\\
Ly$\alpha$	&	0.7	&	1.80e+12	&	1.98e+11	&	63		\\
Ly$\alpha$	&	0.7	&	1.16e+12	&	1.46e+11	&	24		\\
Ly$\alpha$	&	0.7	&	1.01e+12	&	1.33e+11	&	39		\\
Ly$\alpha$	&	0.7	&	9.98e+11	&	5.65e+10	&	40		\\
Ly$\alpha$	&	0.7	&	9.04e+11	&	9.79e+10	&	32		\\
Ly$\alpha$	&	0.7	&	8.40e+11	&	7.55e+10	&	18		\\
										
OVI	&	1.0	&	9.61e+12	&	8.52e+11	&	622		\\
OVI	&	1.0	&	5.85e+12	&	5.50e+11	&	333		\\
OVI	&	1.0	&	2.86e+12	&	2.76e+11	&	141		\\
OVI	&	1.0	&	2.71e+12	&	3.17e+11	&	167		\\
OVI	&	1.0	&	2.39e+12	&	2.28e+11	&	99		\\
OVI	&	1.0	&	1.49e+12	&	1.44e+11	&	92		\\
OVI	&	1.0	&	1.41e+12	&	1.66e+11	&	88		\\
OVI	&	1.0	&	1.28e+12	&	9.45e+10	&	69		\\
OVI	&	1.0	&	1.03e+12	&	9.16e+10	&	53		\\
OVI	&	1.0	&	9.76e+11	&	9.01e+10	&	45		\\
\hline
\end{tabular}
\caption{Properties of the halos that were used as input for the FIREBall IMO.
The first column gives the emission line, the second the redshift.
Column 3 gives the dark matter mass of the halo and column 4 the stellar mass.
The star formation rate is given in column 5.
}
\label{tab:halopropsfireball}
\end{table}

\begin{table}
\centering
\begin{tabular}{ccc}
\hline
input NUV  & predicted SNR  & stack  \\
 AB magnitude &  per resolution element &  (SNR=3) \\
\hline
18	&	3.15	$-$	3.97	&	1	$-$	1	\\
19	&	1.26	$-$	2.51	&	2	$-$	6	\\
20	&	0.50	$-$	1.58	&	4	$-$	36	\\
21	&	0.20	$-$	0.99	&	9	$-$	227	\\
22	&	0.079	$-$	0.63	&	23	$-$	1433	\\
23	&	0.032	$-$	0.40	&	56	$-$	9043	\\
24	&	0.013	$-$	0.25	&	144	$-$	57056	\\
25	&	0.0050	$-$	0.16	&	352	$-$	360000	\\\hline
\end{tabular}
\caption{SNR results for Ly$\alpha$ galaxies with different magnitudes. The first column gives the magnitude, the second the calculated SNR according to equations \ref{eq:snr1} and \ref{eq:snr2}, corresponding to the pessimistic and the optimistic case. The third column gives the number of targets that need to be stacked at the given magnitude in order to reach an SNR of 3 for the two extreme cases.
}
\label{tab:snrresultslya}
\end{table}

\end{document}